%-------------------------
% This paper uses harvmac
%-------------------------
\input harvmac
\noblackbox
%
%
% titlefont
%
%
\edef\tfontsize{ scaled\magstep3}
 \tfontsize  \tfontsize
 \tfontsize \font\titlei=cmmi10 \tfontsize
\font\titleis=cmmi7 \tfontsize \font\titleiss=cmmi5 \tfontsize
\font\titlesy=cmsy10 \tfontsize \font\titlesys=cmsy7 \tfontsize
\font\titlesyss=cmsy5 \tfontsize  \tfontsize
\skewchar\titlei='177 \skewchar\titleis='177 \skewchar\titleiss='177
\skewchar\titlesy='60 \skewchar\titlesys='60 \skewchar\titlesyss='60
%
%\def\titlefont{\def\rm{\fam0\titlerm}% switch to title font
%\textfont0=\titlerm \scriptfont0=\titlerms \scriptscriptfont0=\titlermss
%\textfont1=\titlei \scriptfont1=\titleis \scriptscriptfont1=\titleiss
%\textfont2=\titlesy \scriptfont2=\titlesys \scriptscriptfont2=\titlesyss
%\textfont\itfam=\titleit \def\it{\fam\itfam\titleit}\rm}
%
%
% math symbols
%
%---------------------------------------------------------------------
%
%
%
% space and backspace in l mode
%
\def\lspace{\ifx\answ\bigans{}\else\qquad\fi}
\def\lbspace{\ifx\answ\bigans{}\else\hskip-.2in\fi} % $$\lbspace...$$
%
%
%     curly letters
%
   %curly letters

%
%
%
%     derivatives
%
%
\def\del{\partial}

\def\bar#1{\overline{#1}}

\def\half{{\textstyle{1\over2}}} %puts a small half in a displayed eqn
\def\frac#1#2{{\textstyle{#1\over #2}}} %puts a small fraction
%in a displayed eqn
%
%
%     various math operators
%
%

\def\Im{\mathop{\rm Im}}
\def\Re{\mathop{\rm Re}}

%
%
%
%       relations
%
\def\ltap{\ \raise.3ex\hbox{$<$\kern-.75em\lower1ex\hbox{$\sim$}}\ }
\def\gtap{\ \raise.3ex\hbox{$>$\kern-.75em\lower1ex\hbox{$\sim$}}\ }
\def\gl{\ \raise.5ex\hbox{$>$}\kern-.8em\lower.5ex\hbox{$<$}\ }
\def\roughly#1{\raise.3ex\hbox{$#1$\kern-.75em\lower1ex\hbox{$\sim$}}}
%
%
%       This defines et al., i.e., e.g., cf., etc.
\def\ie{\hbox{\it i.e.}}

\def\frac#1#2{{\textstyle{#1 \over #2}}}

\def\[{\left[}
\def\]{\right]}
\def\({\left(}
\def\){\right)}

\def\rto{\rightarrow}
\def\quar{{1 \over 4}}
\hyphenation{higgs-ino}
\lref\duff{M. Duff, Class. Quantum Grav. {\bf 5} (1988) 189. }
\lref\HL{J. A. Harvey and J. Liu, Phys. Lett. {\bf B268} (1991) 40.}
\lref\banks{T. Banks, M. Dine, H. Dijkstra and W. Fischler,
Phys. Lett. {\bf B212} (1988) 45.}
\lref\gibb{G. W. Gibbons and P. J. Ruback, Commun. Math. Phys. {\bf 115}
(1988) 267.}
\lref\sch{N. S. Manton  and B. J.Schroers, Ann. Phys. {\bf 225} (1993) 290.}
\lref\dab{A. Dabholkar and J. A. Harvey, Phys. Rev. Lett.  {\bf 63} (1989) 719
\semi
A. Dabholkar, G. Gibbons, J. A. Harvey, and F. Ruiz Ruiz,
Nucl. Phys. B {\bf B340} (1990) 33.}
\lref\sena{A. Sen, preprint TIFR/TH/94-03 hep-th/9402002.}
\lref\senb{A. Sen, preprint TIFR-TH-94-08 hep-th/9402032.}
\lref\GNO{P. Goddard, J. Nuyts and D. Olive, Nucl. Phys. {\bf B125}
(1977) 1.}
\lref\WO {E. Witten and D. Olive, Phys. Lett. {\bf 78B} (1978) 97. }
\lref\CSF{ E. Cremmer, J. Scherk and S.  Ferrara,
Phys. Lett. {\bf 74B} (1978) 61.}
\lref\GHL{J. P. Gauntlett, J. A. Harvey and J. Liu,
Nucl. Phys. {\bf B409} (1993) 363.}
\lref\harr{B. Harrington and H. Shepard, Phys. Rev. {\bf D17} (1978) 2122.}
\lref\rossi{P. Rossi, Nucl. Phys. {\bf B149} (1979) 170.}
\lref\hetsol{A. Strominger, Nucl. Phys. {\bf B343} (1990) 167;
E: Nucl. Phys. {\bf B353} (1991) 565.}
\lref\MO{C. Montonen and D. Olive, Phys. Lett. {\bf 72B} (1977) 117. }
\lref\osborn{H. Osborn, Phys. Lett. {\bf 83B} (1979) 321.}
\lref\gpy{D. J. Gross, R. D. Pisarski and L. G. Yaffe, Rev. Mod. Phys.
{\bf 53} (1981) 43.}
\lref\fromzum{B. Zumino, Phys. Lett. {\bf B69}, (1977) 369.}
\lref\don{
J. A. Harvey and A. Strominger, Comm. Math. Phys. {\bf 151} (1993) 221. }
\lref\smon{J. P. Gauntlett,
Nucl. Phys. {\bf B400} (1993) 103; Nucl. Phys. {\bf B411} (1993) 443.}
\lref\blum{J. Blum, preprint EFI-94-04 (hep-th/9401133).}
\lref\grossman{B. Grossman, Phys. Lett. {\bf A61} (1977) 86.}
\lref\dualrefa{A. Font, L. E. Ib\'a\~nez, D. L\"ust and F. Quevedo,
Phys. Lett. {\bf B249} (1990) 35;
S. J. Rey, Phys. Rev. {\bf D43}  (1991) 526.}
\lref\dualrefb{
A. Sen, Nucl. Phys. {\bf B404} (1993) 109; Phys. Lett. {\bf B303} (1993) 22;
Mod. Phys. Lett. {\bf A8} (1993) 2023;
J. H. Schwarz and A. Sen, Nucl. Phys. {\bf B411}
(1994) 35.}
\lref\witten{E. Witten, Nucl. Phys. {\bf B202} (1982) 253.}
\lref\orbi{L. Dixon, J. A. Harvey, C. Vafa and E. Witten, Nucl. Phys. {\bf
B261} (1985) 678;
Nucl. Phys. {\bf B274} (1986) 285. }
\lref\nati{N. Seiberg, Nucl. Phys. {\bf B303} (1988) 286.}
\lref\rohmw{R. Rohm and E. Witten, Ann. Phys. {\bf 170} (1986) 454.}
 \lref\bddf{T. Banks, M. Dine, H. Dijkstra and W. Fischler,
Phys. Lett. {\bf B212} (1988) 45.}
\lref\sgrav{E. Cremmer, J. Scherk and S. Ferrara, Phys. Lett. {\bf 74B} (1978)
61 \semi
E. Bergshoeff, M. de Roo and B. de Wit, Nucl. Phys. {\bf B182} (1981) 173 \semi
M. K. Gaillard and B. Zumino, Nucl. Phys. {\bf B193} (1981) 221 \semi
M. de Roo, Nucl. Phys. {\bf B255} (1985) 515.}
\lref\kkref{R. Sorkin, Phys. Rev. Lett. {\bf 51} (1983) 87 \semi
D. Gross and M. Perry, Nucl. Phys. {\bf B226} (1983) 29.}
\lref\narain{K. S. Narain, Phys. Lett. {\bf B169} (1986) 41.}
\lref\nsw{K. S. Narain, M. H. Sarmadi, and E. Witten, Nucl. Phys. {\bf B279}
(1987) 369.}
\lref\stw{A. Shapere, S. Trivedi and F. Wilczek, Mod. Phys. Lett. {\bf A6}
(1991) 2677.}
\lref\jm{C. V. Johnson and R. C. Myers, preprint hep-th/9406069. }
\lref\atiyah{M. F. Atiyah, ``Geometry of Yang-Mills Fields'' in Lezioni
Fermiane, Accademia Nazionale dei Lincei Scuola Normale Superiore,
Pisa (1979). }
\lref\adhm{M. F. Atiyah, N. J. Hitchin, V. G. Drinfeld and Yu. I. Manin,
Phys. Lett. {\bf 65A} (1978) 425.}
\lref\ah{M.F. Atiyah and N. J. Hitchin, The Geometry and Dynamics of
Magnetic Monopoles, Princeton University Press, 1988.}
\lref\thooft{G. 't Hooft, Nucl. Phys. {\bf B153} (1979) 141 ; Acta Phys.
Austriaca Suppl.
{\bf XXII} (1980) 531.}
\lref\wittwist{E. Witten, Nucl. Phys. {\bf B202} (1982) 253.}
\lref\hofftwo{ G. 't Hooft, Commun. Math. Phys. {\bf 81} (1981) 267.}
\lref\modrefs{This result is implicit in the work of G. 't Hooft, Phys. Rev.
{\bf D14} (1976) 3432,
see also C. Bernard, Phys. Rev. {\bf D19} (1979) 3013. }
\lref\unpub{J. P. Gauntlett and J. A. Harvey, unpublished.}
\lref\nfourref{S. Mandelstam, Nucl. Phys. {\bf B213} (1983) 149 \semi
L. Brink, O. Lindgren and B. Nilsson, Phys. Lett. {\bf 123B}(1983) 323 \semi
P. S. Howe, K. S. Stelle and P. K. Townsend, Nucl. Phys. {\bf B214} (1983) 519
\semi
P. S. Howe, K. S. Stelle and P. K. Townsend, Phys. Lett. {\bf 124B} (1983) 55
\semi
E. Martinec, Phys. Lett. {\bf B171} (1986) 189.}
\lref\wbrane{C. Callan, J. A. Harvey and A. Strominger, Nucl. Phys. {\bf B359}
(1991) 40.}
\lref\dufflu{M. Duff and J. Lu, Nucl. Phys, {\bf B354} (1991) 129; {\bf B357}
(1991) 129.}
\lref\dufflutwo{M. Duff and J. Lu, Phys. Rev. Lett. {\bf 66} (1991) 1402.}
\lref\bmt{P. J. Braam, A. Maciocia, and A. Todorov, Invent. math. {\bf 108}
(1992) 419.}
\lref\kmir{F. Bogomolov and P. J. Braam,
Commun. Math. Phys. {\bf 143} (1992) 641.}
\lref\hoffmod{G. 't Hooft, Phys. Rev. {\bf D14} (1976) 3432.}
\lref\dist{J. Distler and S. Kachru, PUPT-1464, hep-th@xxx/9406091.}
\lref\bhrefs{recent refs on bh, Giddings, Polchinski, Strominger,
Kallosh, Duff???, what else?}
\lref\colea{S. Coleman, ``Classical Lumps and their Quantum Descendants'' in
Aspects of Symmetry, Cambridge University Press, 1985.}
\lref\coleb{S. Coleman, ``The Magnetic Monopole Fifty Years Later,''
in Les Houches 1981, Proceedings, Gauge Theories in High Energy Physics,
Part 1.}
\lref\gorev{P. Goddard and D. I. Olive, ``New Developments in the Theory of
Magnetic
Monopoles,'' Rep. Prog. Phys. {\bf 41} (1978) 1357.}
\lref\presk{J. Preskill, ``Magnetic Monopoles,'' Ann. Rev. of Nucl. and Part.
Sci. {\bf 34}
(1984) 461.}
\lref\diracb{P. Dirac, ``The Theory of Magnetic Poles,'' Phys. Rev. {\bf 74}
(1948) 817.}
\lref\moore{G. Moore, hep-th/9305139.}
\lref\gir{L. Girardello, A. Giveon, M. Porrati, and A. Zaffroni,
hep-th/9406128.}
\lref\cw{S. Coleman and E. Weinberg, Phys. Rev. {\bf D7} (1973) 1888.}
\lref\osbornb{For a more complete discussion see H. Osborn, ``Semiclassical
Methods
for Quantising Monopole Field Configurations,'' in {\it Monopoles in Quantum
Field Theory},
Proceedings of the Monopole Meeting, Trieste Italy 1981,
eds. N. S. Craigie, P. Goddard and W. Nahm (World Scientific, Singapore
(1982)).}
\lref\coleref{ For a  pedagogical discussion see S. Coleman, ``The uses of
Instantons'' in
Aspects of Symmetry, Cambridge University Press, 1985.}
\lref\witteff{E. Witten,  `` Dyons of charge $e \theta/2 \pi $, ''
Phys. Lett. {\bf 86B} (1979) 283. }
\lref\sena{A. Sen, Int. J. Mod. Phys. {\bf A9} (1994) 3707.}
\lref\sendual{A. Sen, Phys. Lett. {\bf 329} (1994) 217.}
\lref\call{C. Callias, `` Index theorems on open spaces,''
Commun. Math. Phys. {\bf 62} (1978) 213.}
\lref\swtwo{N. Seiberg and E. Witten, Nucl. Phys. {\bf B431} (1994) 484.}
\lref\swone{N. Seiberg and E. Witten, Nucl. Phys. {\bf B426} (1994) 19,
Erratum ibid {\bf B430} (1994) 485.}
\lref\segal{G. Segal, unpublished}
\lref\porrati{M. Porrati, ``On the existence of states saturating
the Bogomolny bound in $N=4$ supersymmetry,'' hep-th/9505187}
\lref\gibbman{G. Gibbons and N. Manton, ``Classical and quantum dynamics
of BPS monopoles,'' Ncul. Phys. {\bf B274} (1986) 183.}
\lref\jackson{J. D. Jackson, {\it Classical Electrodynamics} }
\lref\wuyang{T. T. Wu and C. N. Yang, Phys. Rev. {\bf D12} (1975) 3845.}
\lref\dirac{P.A. M. Dirac, Proc. Roy. Soc. {\bf A133} (1931) 60. }
\lref\thmon{G. 't Hooft, Nucl. Phys. {\bf B79} (1974) 276.}
\lref\poly{A.M. Polyakov, JETP Lett. {\bf 20} (1974) 194.}
\lref\rohmwitt{R. Rohm and E. Witten, ``The antisymmetric tensor field
in superstring theory, '' Ann. Phys. {\bf 170} (1986) 454.}
\lref\hl{J. Harvey and J. Liu, ``Magnetic monopoles in $N=4$ supersymmetric
low-energy superstring theory,'' Phys. Lett. {\bf B286} (1991) 40.}
\lref\bogbound{E. B. Bogomol'nyi, Sov. J. Nucl. Phys. {\bf 24} (1976) 449.}
\lref\numref{Some ref to numerical solution of eqns}
\lref\ewein{E. Weinberg, Phys. Rev. {\bf D20} (1979) 936 \semi
E. Corrigan and P. Goddard, Commun. Math. Phys. {\bf 80}
(1981) 575.}
\lref\taubes{C. H. Taubes, Commun. Math. Phys. {\bf 91}
(1983) 235. }
\lref\cardy{J. Cardy and E. Rabinovici, Nucl. Phys. {\bf B205} (1982) 1
\semi J. Cardy, Nucl. Phys. {\bf B205} (1982) 17 \semi A. Shapere
and F. Wilczek, Nucl. Phys. {\bf B320} (1989) 669.}
\lref\sdual{A. Font, L. Ibanez, D. Lust and F. Quevedo, Phys. Lett. {\bf B249}
(1990) 35 \semi S. J. Rey, Phys. Rev. {\bf D43} (1991) 526.}
\lref\corro{E. Corrigan and D. Olive, Nucl. Phys. {\bf B110} (1976) 237.}
\lref\freviews{Various aspects of duality in field theory are reviewed in
K. Intriligator and N. Seiberg, ``Lectures on supersymmetric
gauge theories and electric-magnetic duality,'' hep-th/9509066 \semi
N. Seiberg, ``The power of duality: Exact results in 4-D
susy field theory,'' hep-th/9506077. }
\lref\sreviews{To my knowledge there is no up to date review of duality in
string theory. The following few references might be useful in getting
oriented to some of the earlier work. A. Sen, Int. J. Mod. Phys. {\bf A9}
(1994) 3707 \semi
A. Sen and J. Schwarz, ``Duality symmetries of 4-D heterotic
strings,''
Phys. Lett. {\bf B312} (1993) 105, hep-th/9305185
\semi C. Hull and
P. Townsend, ``Unity of superstring dualities,'' Nucl.
Phys. {\bf B438} (1995) 109, hep-th/9410167 \semi
E. Witten, ``String theory dynamics in various dimensions,''
Nucl. Phys. {\bf B443} (1995) 85, hep-th/9503124 \semi
J. A. Harvey and A. Strominger, ``The heterotic string is a soliton,''
Nucl. Phys. {\bf B449} (1995) 535, hep-th/9504047. }
\lref\fallref{Corrigan and co.? Jaffe and Taubes? }
\lref\gswa{Appendix 5.A of M. Green, J. Schwarz and E. Witten,
{\it Superstring theory: 1}, Cambridge University Press,
Cambridge 1987.}
\lref\wessb{J. Wess and J. Bagger, {\it Supersymmetry and Supergravity},
Princeton University Press, Princeton 1992.}
\lref\jackiwr{R. Jackiw and C. Rebbi, Phys. Rev. {\bf D13} (1976) 3398.}
\lref\gaunharv{J. Gauntlett and J. A. Harvey, ``Duality and the dyon
spectrum in $N=2$ super Yang-Mills theory,'' hep-th/9508156.}
\lref\sethi{S. Sethi, M. Stern and E. Zaslow, ``Monopole and dyon
bound states in $N=2$ supersymmetric Yang-Mills theories,''
hep-th/9508117.}
\lref\mantonmod{N. S. Manton, ``Monopole interactions at long range,''
Phys. Lett. {\bf 154B} (1985) 397.}
\lref\gauntwo{J. P. Gauntlett, ``Low energy dynamics of $N=2$
supersymmetric monopoles,'' Nucl. Phys. {\bf B411} (1994) 443,
hep-th/9305068.}
\lref\blumfour{J. Blum, ``Supersymmetric quantum mechanics of
monopoles in $N=4$ Yang-Mills theory,'' Phys. Lett. {\bf B333} (1994)
92, hep-th/9401133.}
\lref\bss{L. Brink, J. H. Schwarz and J. Scherk, ``Supersymmetric
Yang-Mills theories,'' Nucl. Phys. {\bf B121}
(1977) 77.}
\lref\prasadsomm{M. K. Prasad and C. H. Sommerfield, Phys. Rev. Lett.
{\bf 35} (1975) 760.}
\lref\colest{S. Coleman, Phys. Rev. {\bf D11} (1975) 2088.}
\lref\mand{S. Mandelstam, Phys. Rev. {\bf D11} (1975) 3026.}
\lref\otherbound{refs for string, D-brane bound states.}
\lref\golds{H. Goldstein, {\it Classical Mechanics}, Chapter 6,
Addison Wesley, 1950.}
\lref\vafwit{C. Vafa and E. Witten, ``A strong coupling test of S duality,''
Nucl. Phys. {\bf B431} (1994) 3, hep-th/9408074.}
\lref\hms{J. A. Harvey, G. Moore and A. Strominger, ``Reducing S duality
to T duality,'' Phys. Rev. {\bf D52} (1995) 7161.}
\lref\nfourfinite{M. Sohnius and P. West, Phys. Lett. {\bf 100B}
 (1981) 245 \semi
S.Mandelstam, Nucl. Phys. {\bf B213} (1983) 149 \semi
 P.S. Howe, K.S. Stelle
and P.K. Townsend, Nucl. Phys.{\bf  B214} (1983) 519 \semi
 Nucl. Phys. {\bf B236 }
(1984) 125 \semi  L. Brink, O. Lindgren and B. Nilsson, Nucl. Phys. {\bf B212}
(1983) 401. }
\lref\ntwofinite{P.Howe and K.Stelle and P.West, Phys. Lett. {\bf 124B}
 (1983) 55.}
\lref\wittsusy{E, Witten, Nucl. Phys. {\bf B202} (1982) 253.}
\lref\gibbmannew{G. W. Gibbons and N. S. Manton, ``The moduli space metric
for well separated BPS monopoles,'' Phys. Lett. {\bf B356} (1995)
32, hep-th/9506052.}
\lref\newdual{L. Girardello, A. Giveon, M. Porrati and A. Zaffroni,
Nucl. Phys. {\bf B448} (1995) 127, hep-th/9502057 \semi
O. Aharony and S. Yankielowicz, ``Exact electromagnetic duality
in $N=2$ supersymmetric QCD theories,'' hep-0th/9601011 \semi
J. P. Gauntlett and D. A. Lowe, ``Dyons and S-duality
in $N=4$ supersymmetric gauge theory,'' hep-th/9601085 \semi
K. Lee, E. J. Weinberg and P. Yi ``The moduli space of many BPS monopoles
for arbitrary gauge groups,'' hep-th/9602167 \semi
K. Lee, E. J. Weinberg and P. Yi, ``Electromagnetic duality and
$SU(3)$ monopoles,'' hep-th/9601097.}
\lref\dadda{A. D'Adda, R. Horsley and P. Di Vecchia, Phys. Lett. {\bf 76B}
(1978) 298.}
\lref\oliverev{D. Olive, ``Exact electromagnetic duality,''
hep-th/9508089.}
\lref\ggpz{L. Girardello, A. Giveon, M. Porrati and A. Zaffaroni,
Phys. Lett. {\bf B334} (1994) 331, hep-th/9406128; 
Nucl. Phys. {\bf B448} (1995) 127,
hep-th/9502057.}

%
%-------------------
% title page
%-------------------
%
\Title{\vbox{\baselineskip12pt
\hbox{EFI-96-06}
\hbox{hep-th/9603086}}}
{\vbox{\centerline{Magnetic Monopoles, Duality, and Supersymmetry }
}}
{
\baselineskip=12pt
\bigskip
\centerline{Jeffrey A. Harvey}
\bigskip
\centerline{\sl Enrico Fermi Institute, University of Chicago}
\centerline{\sl 5640 Ellis Avenue, Chicago, IL 60637 }
\centerline{\it Internet: harvey@poincare.uchicago.edu}

\bigskip
\medskip
\centerline{\bf Abstract}
These notes present a pedagogical introduction to magnetic
monopoles and exact electromagnetic duality in supersymmetric
gauge theories. They are based on lectures given at the 1995 Trieste Summer
School in High Energy Physics and Cosmology
and at the 1995 Busstepp Summer School at Cosener's House.
\bigskip
\bigskip

}
\Date{3/96}
%\draftmode
\vfil
\eject
%
%----------------------
% Body of Paper
%----------------------
%\centerline{\bf Contents}
\nobreak\medskip{\baselineskip=12pt
\footnotefont\parskip=0pt\catcode`\@=11
\def\leaderfill#1#2{\leaders\hbox to 1em{\hss.\hss}\hfill%
\ifx\answ\bigans#1\else#2\fi}
\noindent {0.} {Introduction} \leaderfill{2}{2} \par
\noindent \quad{0.1.} {Introduction and Outline} \leaderfill{2}{2} \par
\noindent \quad{0.2.} {Acknowledgements} \leaderfill{4}{4} \par
\noindent \quad{0.3.} {Conventions} \leaderfill{4}{4} \par
\noindent \quad{0.4.} {Exercises} \leaderfill{5}{5} \par
\noindent {1.} {Lecture 1} \leaderfill{5}{5} \par
\noindent \quad{1.1.} {Duality} \leaderfill{5}{5} \par
\noindent \quad{1.2.} {Electromagnetic Duality} \leaderfill{8}{9} \par
\noindent \quad{1.3.} {The Dirac Monopole \'a la Wu-Yang } 
\leaderfill{10}{11} \par
\noindent \quad{1.4.} {The 't Hooft-Polyakov Monopole} \leaderfill{12}{14} \par
\noindent \quad{1.5.} {Exercises for Lecture 1} \leaderfill{17}{19} \par
\noindent {2.} {Lecture 2} \leaderfill{18}{20} \par
\noindent \quad{2.1.} {Symmetric Monopoles and the Bogomol'nyi Bound}
\leaderfill{18}{20} \par
\noindent \quad{2.2.} {The Prasad-Sommerfield Solution} \leaderfill{20}{23}
\par
\noindent \quad{2.3.} {Collective Coordinates and the Monopole Moduli Space}
\leaderfill{21}{24} \par
\noindent \quad{2.4.} {Exercises for Lecture 2} \leaderfill{25}{28} \par
\noindent {3.} {Lecture 3} \leaderfill{25}{29} \par
\noindent \quad{3.1.} {The Witten Effect} \leaderfill{25}{29} \par
\noindent \quad{3.2.} {Montonen-Olive and $SL(2,Z)$-duality} 
\leaderfill{27}{32} \par
\noindent \quad{3.3.} {Exercises for Lecture 3} \leaderfill{30}{35} \par
\noindent {4.} {Lecture 4} \leaderfill{31}{36} \par
\noindent \quad{4.1.} {Monopoles and Fermions} \leaderfill{31}{38} \par
\noindent \quad{4.2.} {Monopoles Coupled to Isospinor Fermions}
\leaderfill{33}{42} \par
\noindent \quad{4.3.} {Monopoles Coupled to Isovector Fermions}
\leaderfill{36}{43} \par
\noindent \quad{4.4.} {Exercises for Lecture 4} \leaderfill{38}{43} \par
\noindent {5.} {Lecture 5} \leaderfill{38}{44} \par
\noindent \quad{5.1.} {Monopoles in $N=2$ Supersymmetric Gauge Theory} \leaderfill{38}{44} \par
\noindent \quad{5.2.} {The Bogomol'nyi bound Revisited} 
\leaderfill{40}{47} \par
\noindent \quad{5.3.} {Monopoles in $N=4$ Supersymmetric Gauge Theory} \leaderfill{42}{48} \par
\noindent \quad{5.4.} {Supersymmetric Quantum Mechanics on ${\cal M}_k$}
\leaderfill{44}{51} \par
\noindent \quad{5.5} {Exercises for Lecture 5} \leaderfill{46}{54} \par
\noindent {6.} {Lecture 6} \leaderfill{47}{54} \par
\noindent \quad{6.1.} {Implications of $S$-duality} \leaderfill{47}{54} \par
\noindent \quad{6.2.} {The Two-monopole Moduli Space From Afar}
\leaderfill{49}{57} \par
\noindent \quad{6.3.} {The Exact Two-monopole Moduli Space} 
\leaderfill{51}{60} \par
\noindent \quad{6.4.} {$S$-duality and Sen's Two-form} 
\leaderfill{53}{62} \par
\noindent \quad{6.5.} {Exercises for Lecture 6} \leaderfill{55}{64} \par
\noindent {7.} {Conclusions and Outlook} \leaderfill{56}{65} \par
\catcode`\@=12\bigbreak\bigskip}

\vfil
\eject

\secno=-1
\newsec{Introduction}
\subsec{Introduction and Outline}

The subject of magnetic monopoles has a remarkable vitality, resurfacing every
few years with new focus. The current interest in magnetic monopoles
centers around the idea of electromagnetic duality.  Exact electromagnetic
duality,
first proposed in modern form by Montonen and Olive \MO, has finally been put
to non-trivial tests \refs{
\sendual, \vafwit, \ggpz, \swtwo, \gaunharv, \sethi } in finite $N=4$ Super Yang-Mills
theory \nfourfinite\
and special
finite $N=2$ theories \ntwofinite.
Although duality is far from being understood, the evidence is now sufficiently
persuasive that the focus has turned from testing to duality to understanding
its consequences and structure. Perhaps more significantly,
it has also been understood that duality plays a central role in understanding
strongly coupled gauge theories with non-trivial
dynamics, particularly in their supersymmetric
form \refs{\swone, \swtwo}.
Here the duality is not exact but nonetheless  the idea of a dual formulation
of a strongly coupled theory in terms of weakly coupled magnetic monopoles is
central and the dynamics of these theories is closely tied to the properties
of magnetic monopoles, many of which can be studied semi-classically.

These lectures are intended to provide an introduction to the
properties of magnetic monopoles which are most relevant to the
study of duality.
They have for the most part been kept at a level which
should be appropriate for graduate students with a good grasp of
quantum field theory and hopefully at least a passing acquaintance  with
supersymmetry  and  some of the
tools of general relativity.  There is also a recent review of
exact electromagnetic duality by
David Olive \oliverev\ which I highly recommend.
There are many topics which are
not covered, including, monopoles in gauge groups other than $SU(2)$,  homotopy
theory as applied to magnetic monopoles, the Callan-Rubakov effect,
astrophysical implications of magnetic
monopoles,  experimental and theoretical bounds on the cosmic monopole
abundance, and in general anything having to do with ``real '' magnetic
monopoles
as they might be found in nature.  These omissions are more than made up
for by the existence of excellent reviews which cover this material
\refs{\gorev,\colea,\coleb,\presk} and which the student should
consult to
complement the present lectures.  I have also not covered most of  the
sophisticated mathematics
related to the structure of the BPS monopole moduli space. A good reference
for this material is \ah.

 What I have tried to do is to take a fairly direct route starting from the
basics of magnetic monopoles and ending at the new evidence for $S$-duality
found by A. Sen in February, 1994 \sendual.  These lectures are organized as
follows.
The first lecture begins with a brief discussion of some early
examples of duality in physical systems.  The basic ideas of electromagnetic
duality and
the Dirac monopole are then introduced followed by a discussion of
the 't Hooft Polyakov monopole.
Lecture two begins a discussion of magnetic monopoles in the BPS limit
of vanishing scalar potential.  The Bogomol'nyi bound is derived and the
BPS single charge monopole solution is presented.  Spontaneous breaking
of dilation symmetry and the Higgs field as ``dilaton'' are discussed.
Collective coordinates are introduced through a concrete
construction of the moduli space of
the charge one BPS solution and then a more formal discussion of the moduli
space
of BPS monopoles is given.  The third lecture discusses the dependence of
monopole
physics on the $\theta$ angle and introduces Montonen-Olive duality and its
generalization to $SL(2,Z)$ known as $S$-duality.
The coupling of fermions to magnetic
monopoles is
explored in the fourth lecture.
Fermion zero modes are constructed and the effects of
their quantization on the monopole spectrum is discussed.
The fifth lecture explores
the consequences of  both $N=2$ and $N=4$ supersymmetry for monopole
physics. The Bogomolnyi bound is revisited and related
to a central extension of
the supersymmetry algebra and the relation between BPS saturated states
and short supermultiplets is briefly discussed.
Finally, the basic features of
supersymmetric quantum mechanics on the monopole moduli
space are presented.  The final
lecture is devoted to the evidence for $S$-duality which comes from an
analysis of the two-monopole moduli space following the work of Sen.
I have included in this lecture a brief explanation of some
elegant work of Manton's on the asymptotics of the two monopole
moduli space. The final section contains some very brief remarks
on open problems and recent developments.
I have also taken the liberty of expanding some
of the lectures beyond the material
actually presented in order to provide what I hope will be a more useful
review of duality.

There have of course been
spectacular new developments in understanding duality in supersymmetric
gauge theories with $N=1,2,4$ supersymmetry and also in understanding
duality in string theory \refs{\freviews, \sreviews} which are
not covered at all in these lectures.
These developments show that electromagnetic duality
is a profound new tool for probing the behavior of strong coupling dynamics.
The material covered here is rather mundane in comparison but hopefully will
provide students with some of the background necessary to appreciate and
contribute
to these new ideas.

\subsec{Acknowledgements}
I would like to thank the organizers of the Trieste Summer School and the
Busstepp
School for the invitations to lecture and the students at these schools for
their
questions and interest.  It is a pleasure to acknowledge
colleagues and collaborators
who have  shared their insights into the topics discussed here.  In particular
I would like to thank
J. Blum, C. Callan, A. Dabholkar, J. Gauntlett, G. Gibbons, J. Liu,
G. Polhemus,
A. Sen, A. Strominger, and E. Witten.

\subsec{Conventions}
We will use standard ``field theory'' relativity conventions with
Minkowski space signature
$(+ - - -)$ and $\epsilon^{0123} = +1$. Greek indices run over the
range
$0,1,2,3$ while Roman indices run over the spatial indices $1,2,3$.

Generators $T^a$ of a compact Lie algebra
are taken to be anti-Hermitian. Gauge fields will be written as
vector fields with an explicit gauge index $A_\mu^a$, as Lie-algebra valued
vector fields, $A_\mu = A_\mu^a T^a$,  or as Lie-algebra valued
one-forms $A = A_\mu dx^\mu$.

The speed of light $c$ will always be set
to $1$. For the most part I will set $\hbar=1$ and denote the gauge coupling by
$e$. We will use Heaviside-Lorentz conventions for electromagnetism
with factors of $4 \pi$ appearing in Coulomb's law rather than
in Maxwell's equations. The electric field of a point charge $q$ is
$$ \vec E = {q \hat r \over 4 \pi r^2}. $$
Similarly the magnetic field far outside a magnetic monopole of magnetic
charge $g$ is given by
$$ \vec B = {g \hat r \over 4 \pi r^2 }. $$
It is common in some monopole literature \coleb\ to use definitions of the
electric
charge $e$  and magnetic charge $g$  which differ by a factor of $4 \pi$ in
order to preserve
the quantization condition in the form originally given by Dirac, $eg=n/2$.
Since
the emphasis of these notes is on duality between electric and magnetic states,
such a convention is inappropriate.  Another common convention in the monopole
literature \gibbman\ leaves $\hbar $ as an independent constant but sets the
gauge coupling
$e=1$.

Other conventions involving supersymmetry and gamma matrices will be
discussed in the text as they arise.

\subsec{Exercises}
Each lecture is followed by a set of exercises. Most of these
are short and straightforward
and are meant to reinforce the material covered rather than to seriously
challenge
the student.  A few of the problems involve somewhat more advanced
topics. As usual, serious students are strongly encouraged to work
most of the problems.
\newsec{Lecture 1}

\subsec{Duality}

Saying that a physical system exhibits ``duality'' implies that
there are two complementary perspectives, formulations, or constructions
of the theory. To begin I will briefly describe duality
in three systems which have had an impact on the search for
duality in four-dimensional gauge theories.

In quantum mechanics we say that there is particle-wave
duality meaning that quantum mechanically particles can exhibit wave like
properties and waves (e.g. light) can exhibit particle like properties.
We can think of this roughly speaking as the relation between the position
space basis of states $\langle x | \psi \rangle $ and the momentum space
basis $\langle p |\psi \rangle $ given by Fourier transform.

The harmonic oscillator provides a simple example of a system
exhibiting ``self-duality'' in the sense that it looks the
same in coordinate space and in momentum space. So consider
the harmonic oscillator Hamiltonian
\eqn\zeroa{ H = {p^2 \over 2m}  + \half m \omega^2 x^2  }
with $[x,p]=i$.
We can define a duality transformation which exchanges
position and momenta by \foot{This is clearly a discrete
subgroup of a continuous
symmetry which rotates $p$ and $x$ into each other.}
\eqn\zerob{ D: x \rightarrow p/m \omega,
\qquad p \rightarrow - m \omega x }
Note that this is a canonical transformation
and thus preserves the commutation relations. Squaring $D$
we find $D^2=P$ with $P$ the parity operator $P: x \rightarrow -x $.
The fact that $D$ is a symmetry of the harmonic oscillator is
reflected in the fact that the ground state wave function and its
Fourier transform are transformed into one another by the action
of $D$, the Fourier transform of a Gaussian wave function is
again Gaussian.
Of course this is a rather trivial system, analogous
to free field theory, and we will see in fact that the duality
here  is closely
related to the electromagnetic duality of free Maxwell theory.
We will later argue for an exact extension of electromagnetic duality
in $N=4$ super Yang-Mills theory which in some poetic sense should
thus be regarded as the harmonic oscillator of four-dimensional
gauge theory.

Another system, also exactly soluble, which exhibits a somewhat
different kind of duality is the Ising model. This is defined
by taking a set of
spins $\sigma_i$ taking the values $\pm 1$ and living on a square
two-dimensional lattice with nearest neighbor ferromagnetic
interactions of strength $J$. The partition function at temperature
$T$ is
\eqn\zeroc{Z(K) = \sum_{\sigma} \exp ( K \sum_{(ij)} \sigma_i \sigma_j )}
where the sum on $i,j$ runs over all nearest neighbors, the sum
on $\sigma$ over all spin configurations, and $K = J/k_B T$.
This theory was solved explicitly by Onsager and exhibits
a first-order phase transition to a ferromagnetic state at a
critical temperature $T_c$. However even before Onsager's solution
the critical temperature was computed by Kramers and Wannier using
duality.  They showed that the partition function \zeroc\ could
be represented in two different ways as a sum over plaquettes of
a lattice. In the first form the sum is over plaquettes of the
original lattice with coupling $K$. In the second form one finds
a sum over plaquettes of the dual lattice (the square lattice whose
vertices are the centers of the
faces of the original lattice ) with coupling
$K^* $ where $\sinh 2K^* = 1/(\sinh 2K)$. Since the dual lattice is
also a square lattice, the two formulations are equivalent, but with
different values of $K$. Note also that high temperature $(K<<1)$ or
weak coupling is mapped to low temperature ($K^* >>1 $) or strong
coupling on the dual lattice. Now if the system is to have a single
phase transition then it must occur at the self-dual point with
$K=K^*$ or $\sinh(2J/k_B T_c) = 1$.

This model provides a more striking example of the use of duality.
Duality provides
non-trivial information about the critical behavior and
relates a strongly coupled theory to a weakly coupled theory.  Since
many of the thorniest problems in theoretical physics involve strong
coupling (e.g. quark confinement, high $T_c$ superconductivity ) it
is very tempting to look for dualities which would allow us to use
a dual weakly coupled formulation to do computations in such
strongly coupled theories.

Another system which adds to this temptation occurs in two-dimensional
relativistic field theory. The sine-Gordon model
is defined by the action
\eqn\zerod{S_{SG} = \int d^2 x \left(
\half \partial_\mu \phi \partial^\mu \phi
+ {\alpha \over \beta^2} \left( \cos \beta \phi -1 \right) \right) . }
This theory has meson excitations of mass $M_m = \sqrt{\alpha}$ and
solitons which
interpolate between different minima of the potential with mass
$M_s =  8 \sqrt{\alpha}/\beta^2$. By expanding the potential to
quartic order we see that $\beta^2$ acts as the coupling
constant for this theory. Thus the soliton mass is large (compared
to the meson mass) at weak coupling.

Remarkably, this theory  is known to be completely equivalent to
an apparently quite different theory of interacting fermions
knows as the Thirring model.  The action of the Thirring model
is
\eqn\zeroe{S_T = \int d^2 x \left( \bar \psi i \gamma_\mu \partial^\mu \psi
+ m \bar \psi \psi -
{g \over 2} \bar \psi \gamma^\mu \psi \bar \psi \gamma_\mu \psi \right) }
At first sight these two theories appear completely different, but
through the miracle of bosonization they are in fact completely
equivalent \refs{\colest, \mand}.
The map  between the two theories
relates the couplings through
\eqn\zerof{{\beta^2 \over 4 \pi} = {1 \over 1 + g/\pi } }
and maps the soliton of the SG theory to the fundamental fermion
of the Thirring model and the meson states of the SG theory
to fermion anti-fermion bound states. As in the Ising model,
we see from \zerof\ that strong coupling in one theory (i.e. large
$g$) is mapped to weak coupling (small $\beta $ ) in the other theory.
Thus duality provides a means of performing strong coupling
calculations in one theory by mapping them to weak coupling
calculations in a dual theory.

{}From these example we can extract certain general features of
duality symmetries, although not all may be present in all
examples. First, duality relates weak and strong coupling.
Second, it interchanges fundamental quanta with solitons and
thus exchanges Noether charges with topological charges. Finally
it often involves a geometric duality, for example
relating lattices
to their duals.  In four dimensional supersymmetric gauge theories
we will find obvious generalizations of the first two features.
The geometrical aspects of duality are also present, but only
become clear when one considers general gauge groups.

The search for duality in four-dimensional gauge theories seems
to have been motivated by the existence of dualities in
these simpler systems, by electromagnetic duality and the results of
Dirac, 't Hooft and Polyakov regarding the possible existence
of magnetic monopoles, and by the work of Mandelstam, 't Hooft and
others suggesting that confinement in QCD might arise as a
dual form of superconductivity involving condensation of some
sort of magnetically charged objects.

In spite of these hints, it has only been in the
last two years  that the idea
of duality in non-trivial four-dimensional theories has been
taken seriously by most particle physicists. These lectures will
lead up to one non-trivial test of duality, but the skeptic
could certainly remain unconvinced by the evidence
discussed here. Although the conceptual
underpinnings of duality remain quite mysterious,  recent
developments in gauge theory and string theory leave little room
to doubt that duality exists and has significant applications.

\subsec{Electromagnetic Duality}

Maxwell's equations  read
\eqn\onea{\eqalign{ \vec \nabla \cdot \vec E  & = \rho_e \qquad \qquad \vec
\nabla \cdot
\vec B = 0 \cr
\vec \nabla \times \vec B - {\partial \vec E \over \partial t } & = \vec J_e
\qquad \qquad
\vec \nabla \times \vec E + {\partial \vec B \over \partial t} = 0 . \cr }}
When  $\rho_e = {\vec J}_e = 0$ these equations are invariant under
the duality transformation
\eqn\oneb{D: \quad \vec E \rightarrow \vec B,  \qquad \vec B \rightarrow - \vec
E .}
Note that $D^2$ takes $(\vec E, \vec B) \rto  (- \vec E, - \vec B )$ which is a
transformation
by charge conjugation, $C$ \foot{Note the analogy with the duality
transformation \zerob\ for the harmonic oscillator. This analogy can
be made precise by decomposing the electromagnetic field
in terms of normal modes.}.  Thus the first thing we learn is that theories
with exact duality must also be invariant under charge conjugation. The duality
transformation \oneb\ can be generalized to
duality rotations parameterized by
an arbitrary angle $\theta$,
\eqn\onec{\eqalign{  \vec E & \rightarrow \cos \theta \vec E +
\sin \theta \vec B, \cr
\vec B & \rightarrow  - \sin \theta \vec E + \cos \theta \vec B .\cr }}
We will see later that this continuous duality transformation is broken to a
discrete
subgroup by instanton effects when duality is embedded in non-abelian gauge
theories.
If we write the Maxwell equations in covariant form in terms of the
field strength $F^{\mu \nu}$ with $F^{0i} = - E^i$ and
$F^{ij} = -\epsilon^{ijk} B^k $
then we have
\eqn\oned{ \partial_\mu F^{\mu \nu} = j_e^\nu, \qquad \partial_\mu *F^{\mu \nu}
= 0 }
where $* F^{\mu \nu} = \half \epsilon^{\mu \nu \lambda \rho}F_{\lambda \rho}$
and
the duality transformation \oneb\ takes the form $F_{\mu \nu} \rightarrow
*F_{\mu \nu}$. Note that in Minkowski space $** = -1$ in agreement with
$D^2 = -1$.

The duality symmetry of the free Maxwell equations is broken by the presence
of electric source terms. For this reason it is of no practical interest in
everyday
applications of electromagnetism.  However the possibility that such a symmetry
might nonetheless exist in some more subtle form has long intrigued physicists.

If we are to have such a symmetry it is clear that we will have to make the
equations \onea\ symmetric  by including magnetic
source terms so that $\partial_\mu *F^{\mu \nu} = k^\nu$ with $k^\nu$ the
magnetic
four-current.  Of course in standard electromagnetism we actually take
advantage of
the lack of such source terms to introduce a vector potential.  That is using
$\partial_\mu *F^{\mu \nu}=0$ we  write $F^{\mu \nu} = \partial^\mu A^\nu -
\partial^\nu A^\mu$  with $A^\mu = (\Phi, \vec A)$
the vector potential. The association of a vector potential
to a given field strength is not unique. The ambiguity is that of gauge
transformations
\eqn\onee{A_\mu \rightarrow A_\mu - \partial_\mu \chi }
which leave the field strength invariant.

Now recall that in coupling electromagnetism to quantum mechanics it is
the vector potential $A_\mu$ and not just the field strength that plays a
central role.  Minimal coupling involves replacing the momentum operator
by its covariant generalization
\eqn\onef{ \vec p = -i \vec \nabla \rto -i (\vec \nabla - i e \vec A) }
with $e$ the electric charge.  The Schr\"odinger equation
\eqn\oneg{ i {\partial \psi \over \partial t} = - {1 \over 2 m} (\vec \nabla
-ie \vec A)^2 \psi + V \psi}
is then invariant under the combination of a gauge transformation on the
vector potential and a phase transformation of the wave function:
\eqn\oneh{\eqalign{ \psi & \rightarrow e^{-i e \chi} \psi \cr
                                       \vec A & \rightarrow \vec A - \vec
\nabla \chi \equiv \vec A - {i \over e}
                                        e^{i e \chi} \vec \nabla e^{-i e \chi}
.  \cr }}
The latter form of the gauge transformation has been used to indicate that
the fundamental quantity is the $U(1)$ group element $e^{i e \chi}$ and not
$\chi$ itself.

Now returning to duality, we can ask, following Dirac, whether it is possible
to
add magnetic source terms to the Maxwell equations without disturbing the
consistency
of the coupling of electromagnetism to quantum mechanics.  Dirac's argument
\dirac, adapted
to a modern perspective following the work of Wu and Yang \wuyang\ is given in
the following section.

\subsec{The Dirac Monopole \'a la Wu-Yang}

Since we certainly do not believe that electromagnetism  is correct down to
arbitrarily small distance scales, let us first try to find a consistent
description of
a magnetic monopole excluding from consideration a region of radius $r_0$
around
the center of the monopole. That is for $r>r_0$ we have a magnetic field
\eqn\onei{ \vec B = {g \hat r \over 4 \pi r^2 } }
and we want to find a consistent description of quantum mechanics for
$r> r_0$ in the presence
of such a monopole magnetic field.  Mathematically we are looking for a
description
in $R^3 - \{ 0 \}$.

To couple a  quantum mechanical charged particle to a
background field we need the vector potential, but
this seems inconsistent with having a  magnetic monopole field.
The solution involves
 making use of the ambiguity relating the vector potential
to the field strength.
To be specific, we can try to use different vector potentials in different
regions
as long as the difference between them on overlap regions is
that of a gauge
transformation. Then the physically measurable field strength will be
continuous
and well defined. The simplest way to accomplish  this  is to divide a
two-sphere $S^2$
of fixed radius $r>r_0$ into a Northern half $N$ with $0 \le \theta \le \pi/2$,
a Southern half $S$ with $\pi/2 \le \theta \le \pi $ and the overlap region
which
is the equator $E$ at $\theta=\pi/2$ (if desired the overlap region can be
taken
to be a band of finite width including the equator).  The vector potential on
the two halves is then taken to be \wuyang\
\eqn\onej{\eqalign{ {\vec A}_N & = {g \over 4 \pi r} {(1-\cos \theta) \over
\sin \theta }
{\hat e}_\phi \cr
{ \vec A}_S & = - {g \over 4 \pi r} {(1+\cos \theta) \over \sin \theta }
 {\hat e}_\phi . \cr }}
Note that on the two halves of the two-sphere the magnetic field as given
by $\vec B = \vec \nabla \times \vec A$ agrees with \onei.  Note also that
$A_{N,S}$
have singularities  on $(S,N)$  but are well defined in their
respective patches.

Now to see if this construction makes sense we must check that the difference
between $A_N$ and $A_S$ on the overlap region is indeed a gauge transformation.
We have at $\theta= \pi/2$
\eqn\onek{ {\vec A}_N - {\vec A}_S = - \vec \nabla \chi, \qquad
\chi = -{g \over 2 \pi} \phi , }
so that the difference is a gauge transformation. However   the gauge function
$\chi$ is not continuous. This was in fact inevitable as the following
calculation of
the enclosed magnetic charge demonstrates:
\eqn\onel{g = \int_N { \vec B}_N \cdot d \vec S + \int_S {\vec B}_S \cdot d
\vec S =
\int_E ( {\vec A}_N - {\vec A}_S )  \cdot d \vec l = \chi(0) - \chi (2 \pi). }
But physics does not require that $\chi$ be continuous. As is clear from
\oneh, physical quantities will be continuous as long as $e^{-i e \chi}$ is
continuous.  This then gives us the condition $e^{-i e g}=1$ or
\eqn\onem{ eg = 2 \pi n, \qquad n \in Z }
which is the celebrated Dirac quantization condition \dirac.

Let me pause to make a few comments about what we have done so far.

\item{1.} As observed by Dirac, the presence of a single magnetic monopole
anywhere
in the universe is sufficient to guarantee that electric charge must be
quantized. The quantization
of electric charge is of course one of the fundamental experimental facts in
particle physics
and this provides an attractive explanation of why this should be the case.
\item{2.} The $U(1)$ group of gauge transformations has elements $e^{-ie
\chi}$. If
charge is quantized in units of some fundamental quanta $e_1$ then $\chi=0$ and
$\chi= 2 \pi /e_1$ give the same gauge transformation. That is the range of the
parameter
$\chi$ is compact. It is useful to make the distinction between the compact
one-parameter
group which we will call $U(1)$ and the non-compact one-parameter group which
we will call $R$ which arises if charge is not quantized and thus the parameter
range
is the whole real line.  Magnetic monopoles require a compact
$U(1)$ gauge group. Conversely, whenever a theory has a compact $U(1)$
gauge group it has magnetic monopoles. As we will see this includes grand
unified theories where the $U(1)$ group is compact because it is embedded
in a compact Lie Group such as $SU(2)$ but it also includes Kaluza-Klein
theory where the $U(1)$ arises from symmetries of a compact space and
string theory where the compactness seems necessary but is not
yet completely understood in all cases.
\item{3.} Mathematically what we have done is to construct a non-trivial $U(1)$
principal fibre bundle. The base manifold is an $S^2$ at fixed radius which
we cover with two coordinate patches. The fibers are elements of $U(1)$.
The fibers are patched together with gauge transformations which are the
transition functions. The magnetic charge is the first Chern class of this
principal fibre bundle.
\item{4.} In the real world the low-energy gauge group includes more that just
the $U(1)$ of electromagnetism, it includes $SU(3)$ of color and there are
quarks
which carry fractional electric charge. Must the Dirac condition be satisfied
with
respect to the electron or the quarks? The answer to this involves delving into
monopoles in grand unified gauge theories. The brief answer is that,
viewed from a distance,
color is confined and monopoles must only satisfy the Dirac condition with
respect
to the electron.  When viewed close-up such monopoles must also carry a color
magnetic charge and the combination of the color magnetic charge
and ordinary magnetic charge must satisfy a generalization of the
Dirac quantization condition. For details see \corro.

\subsec{The 't Hooft-Polyakov Monopole}
So far we have argued that there is a sensible quantum mechanics which
includes magnetic monopoles as long as the charge is quantized and as
long as we do not ask what happens inside the monopole.  However
there is in this framework no way to determine most of the properties
of these monopoles including their mass, spin, and other quantum
numbers. I now want to discuss
a beautiful result of 't Hooft and Polyakov \refs{\thmon, \poly} which allows
us to probe inside
the monopole and study its properties in detail.

Given that monopoles make sense if and only if the $U(1)$ gauge group is
compact, it makes sense to look for them in theories where $U(1)$ is compact
because it is embedded inside a larger compact gauge group \foot{Compact $U(1)$
groups can also arise in Kaluza-Klein theory  and in string theory
and there one also finds magnetic monopole
solutions \refs{\kkref, \rohmwitt, \hl}. }. The simplest
possibility is the embedding $U(1) \subset SU(2)$ and it is this possibility
which
will occupy our attention for most of these lectures.  We will take as a
starting
point the Yang-Mills-Higgs Lagrangian
\eqn\onen{ {\cal L} = - { 1 \over 4} F_{\mu \nu}^a F^{a \mu \nu} + {1 \over 2}
D^\mu \Phi^a D_\mu \Phi^a - V(\Phi) }
where
\eqn\onena{F_{\mu \nu}^a = \partial_\mu A_\nu^a - \partial_\nu A_\mu^a + e
\epsilon^{abc}
A_\mu^b A_\nu^c }
and the covariant derivative of $\Phi$ is
\eqn\onenb{ D_\mu \Phi^a = \partial_\mu \Phi^a + e \epsilon^{abc} A_\mu^b
\Phi^c }
with $a,b,c=1,2,3$
labeling the adjoint representation of $SU(2)$.
The potential $V(\Phi)$ is chosen so that the vacuum  expectation value of
$\Phi$ is
non-zero.  To be concrete we take $V(\Phi) = \lambda (\Phi^a \Phi^a - v^2)^2
/4$.

By varying \onen\ with respect to $A_\mu^a$ and $\Phi^a$ we obtain the
equations of motion
\eqn\oneo{\eqalign{ D_\mu F^{a \mu \nu} & = e \epsilon^{abc} \Phi^b D^\nu
\Phi^c \cr
                                     (D^\mu D_\mu \Phi)^a & = -\lambda \Phi^a
(\Phi^b \Phi^b - v^2) . \cr }}
The Bianchi identity,
\eqn\onep{ D_\mu *F^{a \mu \nu} = 0 , }
follows from the definition of $F^{a \mu \nu}$.

We will for the most part be interested in static solutions to the equations of
motion
\oneo. We will later include quantum effects by quantizing small fluctuations
about such classical solutions.

It will also be useful in what follows to have an expression for
the energy-momentum tensor for this theory. Straightforward computation
gives \foot{A canonical computation is tedious and requires the addition
of improvement terms to obtain the symmetric and gauge invariant
answer given here. A better procedure is to couple the theory to
a background metric $g_{\mu \nu}$ and to define the energy-momentum
tensor as the variation of the action with respect to the background
metric.}
\eqn\onepa{\Theta^{\mu \nu} =  -F^{a \mu \rho} F_\rho^{a \nu} +
 D^\mu \Phi^a D^\nu \Phi^a - \eta^{\mu \nu} {\cal L} }
For $v=0$, or for vanishing potential ( $\lambda=0$ ) the theory
defined by \onen\
has a classical scale symmetry\foot{The theory is also conformally
invariant but this will not play an important role in what
follows, basically because spontaneous breaking of scale and
conformal invariance only leads to a single Nambu-Goldstone
boson.}. The conserved current is
the dilation current $D_\mu = x^\nu \Theta_{\mu \nu} $ with
\eqn\onepb{\partial_\mu D^\mu = \Theta^\mu_\mu = 0 . }
The case $V(\Phi) \equiv 0 $ will occupy us later. We will argue
that it still makes sense in this case to choose $\Phi$ to have
an arbitrary but non-zero expectation value. This choice spontaneously
breaks scale invariance. The resulting Nambu-Goldstone boson is
traditionally called the dilaton, and is not to be confused with
the ``dilaton'' field in string theory. Quantum mechanically scale
invariance is broken by renormalization and the trace of the
energy momentum tensor is proportional to the beta function of the
theory.  Finite theories with vanishing beta
function can thus exhibit quantum scale invariance.  The simplest example
of this phenomenon occurs in $N=4$ supersymmetric Yang-Mills theory.
Understanding the monopole spectrum in these theories is one
of the goals of these lectures. We will return later to
the implications of spontaneously broken scale invariance.

For now we proceed with a discussion of the theory with non-zero
potential. We want to discuss non-trivial solutions to the classical
equations of motion but before doing that it will be useful to
first discuss the vacuum structure
of the theory. The energy density of any field configuration is given by the
$(0,0)$
component of the energy-momentum tensor,
\eqn\oneq{ \Theta_{00} = \half \left( \vec {E^a} \cdot \vec {E^a} + \vec {B^a}
\cdot \vec {B^a}
+ \Pi^a \Pi^a + \vec D \Phi^a \cdot \vec D \Phi^a \right) + V(\Phi) }
where $\Pi^a$ is the momentum conjugate to $\Phi$, $\Pi^a = D_0 \Phi^a$ and
$E^a$ and $B^b$ are the non-abelian electric and magnetic fields,
\eqn\oner{\eqalign{  E^{ai} & = - F^{a0i} \cr B^{ai} & = - \half \epsilon^{ijk}
F_{jk}^a \cr }}
It is clear that $\Theta_{00} \ge 0$ with equality if only if $F^{a \mu \nu} =
D^\mu \Phi^a =
V(\Phi) = 0$. The vacuum is thus given by a configuration with vanishing gauge
field
and with a constant Higgs field $\Phi^a$ with $\Phi^a \Phi^a = v^2$.
\foot{ I will abuse notation
by writing $\Phi^a$ for the vacuum expectation value of the operator $\Phi^a$
and hope that
it will be clear from the context when I am discussing the full field and when
I am discussing
only its vacuum expectation value.}.  A constant Higgs field breaks the gauge
symmetry
from $SU(2)$ down to a $U(1)$ subgroup \foot{This of course is also an abuse of
terminology but
one that is universal. Gauge symmetries are not true symmetries but are
redundancies
in our description of the configuration space of the theory. As such they are
never
broken.}.  The perturbative spectrum consists of a massless photon,
massive spin one gauge bosons $W^\pm$ with mass $ev $ and a Higgs field
with a mass depending on the second derivative of the potential $V$
at its minimum. For the previous choice of potential the mass is
$m_H  = \sqrt{2 \lambda} v $.

We can define the Higgs vacuum to be the set of all Higgs configurations which
minimize the potential,
\eqn\ones{  M_H = \{ \Phi : V( \Phi ) = 0 \} }
In our example this space is just the two-sphere given by $\sum_a \Phi^a \Phi^a
= v^2$.

So far we have considered the vacuum configuration and the perturbative
excitations about this vacuum.
Now finite energy configurations need not lie in the Higgs vacuum everywhere
but
they must lie in $M_H$ at spatial infinity.
Thus for a finite energy configuration the Higgs field $\Phi^a$, evaluated as
$r \rto \infty$,
provides a map from the $S^2$ at spatial infinity into the $S^2$ of the Higgs
vacuum,
\eqn\onet{ \Phi: S^2_\infty \rto M_H = S^2 .}
Such maps are characterized by an integer which measures the winding of
one $S^2$ around the other (see Exercise 3).  Mathematically,  the
second homotopy group of $S^2$ is the integers,  $\pi_2(S^2) = Z$.

We have argued so far that finite energy configuration have a topological
classification
and that the gauge symmetry of $SU(2)$ is broken down to $U(1)$. What is the
connection with magnetic monopoles?
I will first give a hand-waving argument. Consider a Higgs field configuration
$\Phi^a$
with winding $N \ne 0$. If the gauge field $A_\mu^a$ vanishes then we have for
the total energy
\eqn\oneu{{\rm Energy} = \int d^3 x \half \vec \nabla \Phi^a \vec \nabla \Phi^a
+ \half  \dot {\Phi^a} \dot {\Phi^a} + V(\Phi) \ge
 \int d^3 x \half \vec \nabla \Phi^a \vec \nabla \Phi^a .}
Now write the gradient term as a radial derivative plus an angular derivative
\eqn\onev{ (\vec \nabla \Phi^a)^2 = \left( { \partial \Phi^a \over \partial r}
\right)^2 +
(\hat r \times \vec \nabla \Phi^a )^2 }
If $N \ne 0$ then there must be non-vanishing angular derivatives of $\Phi^a$
at infinity which make the second term in \onev\ go like $r^{-2}$ for large
$r$.
Therefore the total energy is
\eqn\onew{ {\rm  Energy}  > \int {r^2 dr \over r^2} }
which diverges linearly. Therefore to have finite energy configurations with
$N \ne 0$ we must have non-zero gauge fields. From the above argument it is
clear what we need to ensure finite energy. With non-zero gauge fields the
energy involves the covariant derivative of $\Phi^a$ rather than the
ordinary derivative,  we can make the energy finite if there is a cancellation
between the angular part of the vector potential  (which must therefore fall
off
as $1/r$) and the angular derivative of $\Phi$.  This $1/r$ falloff in the
angular component
of $A_\mu$ gives rise to a non-zero magnetic field at  infinity. Thus the
connection
between topology and magnetic charge comes about by demanding finite
energy of the field configuration.

This hand-waving argument can be made somewhat more explicit. I will
however not try to make it less hand-waving. The reader who wishes
a more precise argument should consult the monopole reviews
cited earlier.
In order to have finite energy we need
to ensure that the covariant derivative of $\Phi^a$ falls off faster than $1/r$
at infinity.  Let us write
\eqn\onex{ D_\mu \Phi^a = \partial_\mu \Phi^a + e \epsilon^{abc} A_\mu ^b
\Phi^c \sim 0}
to indicate that the leading $1/r$ terms must vanish at large $r$. Then the
general solution
for the gauge field, to this order, is given by
\eqn\oney{ A_\mu^a \sim  - {1 \over e v^2} \epsilon^{abc} \Phi^b \partial_\mu
\Phi^c +
{1 \over v} \Phi^a A_\mu }
with $A_\mu$ arbitrary.

If we now compute the leading order behavior of the non-abelian gauge
field we find
\eqn\onez{ F^{a \mu \nu} = {1 \over v} \Phi^a  F^{\mu \nu} }
with
\eqn\oneaa{ F^{\mu \nu} = -{1 \over e v^3} \epsilon^{abc} \Phi^a \partial^\mu
\Phi^b
\partial^\nu \Phi^c + \partial^\mu A^\nu - \partial^\nu A_\mu }
and the equations of motion imply $\partial_\mu F^{\mu \nu} = \partial_\mu
*F^{\mu \nu}=0$.
Thus we learn that outside the core of the monopole the non-abelian gauge field
is purely in the direction of $\Phi^a$, that is the direction of the unbroken
$U(1)$.
The magnetic charge of this field configuration is then
\eqn\onebb{ g = \int_{S^2_\infty} \vec B \cdot d \vec S = {1 \over 2 e v^3}
\int_{S^2_\infty}
\epsilon^{ijk} \epsilon^{abc}
\Phi^a \partial^j \Phi^b \partial^k \Phi^c = {4 \pi N \over e} }
with $N$ the winding number of the Higgs field configuration.

We thus find  a quantization condition
\eqn\onecc{ eg = 4 \pi N.}
This is same as the Dirac quantization condition \onem\ for even values of $n$
in \onem. The reason for the additional restriction on $n$
 is that in this theory we could add fields in the
fundamental ${\bf 2}$ representation of $SU(2)$. These would carry electric
charge
$\pm e/2$ and the Dirac quantization condition with regard to these charges
requires \onecc.

It may seem rather puzzling that we have found the same quantization
method by rather different arguments. The Dirac argument just relies
on what goes on outside the monopole while the quantization condition
for 't Hooft Polyakov monopoles involved the topology of non-Abelian
Higgs fields. The connection between these two points of view
is beautiful and deep and is covered in  detail in
\refs{\gorev, \colea}. Very briefly, if we are given
a gauge group $G$ broken down to a subgroup $H$ by the Higgs field
expectation value then the vacuum manifold is $M_H = G/H $
(modulo some technical assumptions) and the topology of the
Higgs field at infinity is classified by $\pi_2(G/H)$. On the
other hand, at infinity the gauge group is just $H$, and the
Wu-Yang version of the Dirac monopole involves patching $H$ gauge
fields along the equator and these are classified by $\pi_1(H)$.
 But a famous
result (proved in the references) asserts that
\eqn\onedd{\pi_2(G/H) = \pi_1(H) }
(as long as $G$ is simply connected) thus providing the link
between the two points of view.

\subsec{Exercises for Lecture 1}
\item{E1.} Consider a point electric charge $e$ and a point magnetic charge
$g$.
Compute the field angular momentum
\eqn\exone{\vec L = \int d^3r \vec r \times (\vec E \times \vec B ) \qquad }
and
\itemitem{a)} Show that $\vec L$ is well-defined and independent of the
distance
between $e$ and $g$.
\itemitem{b)} Show that demanding that the angular momentum be quantized in
units
of $\hbar/2$ yields the Dirac quantization condition.
\item{E2.}
\itemitem{a)}By generalizing Exercise 1 or otherwise prove the
Dirac-Zwanziger-Schwinger
condition for two point charges (dyons)  with combined (electric,magnetic)
charges $(e_1,g_1)$
and $(e_2,g_2)$
\eqn\extwo{ e_1 g_2 - e_2 g_1 = 2 \pi n .}
\itemitem{b)} Explore the allowed solutions to \extwo\ assuming the existence
of
an electron with charges $(e,0)$. Recall that under CP $(e,g) \rto (-e,g)$.
Show that there are solutions which lead to a
CP violating dyon spectrum.  Show that there are $CP$ invariant
solutions with dyons carrying
half of the electric charge of an electron.
\item{E3.}
\itemitem{a)} If $ \Phi^a \rto v \hat {r^a} $ as $ r \rto \infty$ show that
\eqn\exthree{ N = {1 \over 8 \pi v^3} \int_{S^2_\infty} d S^i \epsilon^{ijk}
\epsilon^{abc}
\Phi^a \partial^j \Phi^b \partial^k \Phi^c = 1. }
\itemitem{b)} Construct a map  $S^2_\infty \rto S^2$ having arbitrary integer
winding number $N$.

\newsec{Lecture 2}

\subsec{Symmetric Monopoles and the Bogomol'nyi Bound}

In the previous lecture we have argued that finite energy configurations with
non-zero
topological charge in the theory defined by \onen\
 are necessarily magnetic monopoles satisfying the Dirac
quantization condition.  While one can argue indirectly for the existence
of such solutions to the equations of motion,  it would be nice to construct
solutions directly. Unfortunately,
in general this only turns out to be possible
numerically.

To construct a solution even numerically it is necessary to make
some simplifying assumptions regarding the form of
the gauge and Higgs fields. We would expect
the lowest energy solution to be the one of highest symmetry
compatible with having non-zero topological charge. The theory defined
by \onen\ is Lorentz invariant and hence rotationally invariant.
Let $J^i$ be the generators of the rotation group $SO(3)_R$. Since the
scalar Higgs field must vary at infinity to have non-zero topological
charge it is clear that the solution cannot be invariant
under $SO(3)_R$. The Lagrangian \onen\ also is invariant under
global gauge transformations by the group $SO(3)_G$ with
generators $T^a$. Since the vacuum expectation value of
the Higgs field is non-zero the monopole solution
 cannot be invariant under $SO(3)_G$.
However, it is allowed to be invariant under
the $SO(3)$ diagonal subgroup of the product of
rotations and global gauge transformations $SO(3)_R \times SO(3)_G$, that
is it is invariant under the generators $\vec K = \vec J + \vec T $.
By imposing this $SO(3)$ symmetry as well as a $Z_2$ symmetry
which consists of parity plus a change of sign of $\Phi$ one is left
with a fairly simple ansatz in terms of two radial functions $H$, $K$:
\eqn\blatz{\eqalign{ \Phi^a &= {\hat r^a \over e r} H(ver) \cr
                                      A_i^a&  = - {\epsilon^a}_{ij} {\hat r^j
\over er} (1 - K(ver)) . \cr }}

Substituting this ansatz into the equations of motion \oneo\ yields coupled
differential equations for $H,K$ which can be solved numerically subject
to the boundary conditions
\eqn\blatza{\eqalign{K(ver) & \rto  1, \quad H(ver) \rto 0,
\qquad r \rto 0; \cr
 K(ver) & \rto  0, \quad H(ver)/(ver) \rto 1,
\qquad r \rto \infty . \cr }}

In these lectures we will not need the detailed form
of these solutions and for the most part will be interested in a specific limit
of the equations \oneo\ where an explicit solution is available.  To understand
the nature of this limit we first discuss a general bound on the mass of
configurations with non-zero winding number known as the Bogomol'nyi bound
\bogbound.

To prove the Bogomol'nyi bound we first note that we can write the magnetic
charge as
\eqn\twoa{ g = \int_{S^2_\infty} \vec B \cdot d \vec S = {1 \over v}
\int_{S^2_\infty}
\Phi^a \vec {B^a} \cdot d \vec S = {1 \over v} \int \vec {B^a} \cdot (\vec D
\Phi)^a d^3 r }
using the Bianchi identity $\vec D \cdot \vec {B^a} =0$ and integration by
parts.
Then if we consider a static configuration with vanishing electric field the
energy (mass)
of the configuration is given by
\eqn\twob{\eqalign{ M_M & = \int d^3 r \left( \half ( \vec {B^a} \cdot \vec
{B^a} + \vec D \Phi^a
\cdot \vec D \Phi^a)  + V(\Phi)  \right) \ge \int d^3 r \half ( \vec {B^a}
\cdot \vec {B^a} + \vec D \Phi^a
\cdot \vec D \Phi^a) \cr
&= \half \int d^3 r ( \vec {B^a} - \vec D \Phi^a) \cdot ( \vec {B^a} - \vec D
\Phi^a) + vg }}
using \twoa. We thus have the bound
\eqn\twoc{M_M \ge vg }
with equality iff $V(\Phi) \equiv 0$
and the first-order Bogomol'nyi equation
\eqn\twod{ \vec {B^a} = \vec D \Phi^a }
is satisfied. Note that the bound has been derived for positive magnetic
charge. For negative
magnetic charge we get \twoc\ with  a minus sign on the right. Thus the general
bound is $M_M \ge |vg| $.  In Exercise 4 this bound is generalized to include
configurations
with non-zero electric field as well.

To saturate the bound \twoc\ we require that the potential vanish identically
and
that the Bogomol'nyi equation \twod\ be satisfied.  Let us first discuss
vanishing potential.
Classically we are free to choose $V(\Phi)=0$ but we know that quantum
mechanically
there will be corrections to $V$ \cw.  Eventually we will consider
supersymmetric
theories which have potentials with exact flat directions protected by
supersymmetry.
For the meantime we will consider just the classical theory and impose $V(\Phi)
\equiv 0$
by hand. The next question is whether symmetry breaking makes sense with
$V(\Phi) \equiv 0$.
We can impose as a boundary condition that $\Phi^a \Phi^a \rto v^2$ as $r \rto
\infty$ for
arbitrary $v$. Although there is no potential, a change of the theory from one
value of
$v$ to another value requires changing the Higgs field at infinity. Since we
are in
infinite volume such a motion requires infinite action, even in the absence of
a potential.
Therefore for each value of $v$ the imposition of  this boundary condition at
infinity
gives a well defined Hilbert Space which does not mix with Hilbert spaces built
on
other values of $v$. In other words each value of $v$ determines a
superselection
sector of the theory.

\subsec{The Prasad-Sommerfield Solution}
Following the previous discussion we now proceed to look for a solution
of \twod\ with
spherical symmetry.
The ansatz \blatz\  when substituted
into \twod\  yields the equations
\eqn\twoe{ y K' = -KH;  \qquad  y H' = H - (K^2 -1) }
with $y= ver$ and $H' = dH/dy$.   Manipulation of these equations yields
the solution \prasadsomm
\eqn\twof{\eqalign{ H(y)&  = y \coth y -1 \cr  K(y)
& = {y \over \sinh y }. \cr }}

The long range behavior of this solution is important. At large
$r$,  $K$ vanishes exponentially at distances greater than
$1/(ev) = 1/M_W$
with $M_W$ the mass of the $W^\pm$ gauge bosons resulting from
spontaneous symmetry breaking. Physically this means that there are
$W^\pm$ fields excited in the core of the monopole, but that outside
the core the magnetic field falls like $1/r^2$ as required for a magnetic
monopole.  The form of the Higgs field is also interesting. There is
an exponentially decaying piece, but also a piece which falls of only
as $1/r$. For large $r$ we have
\eqn\twofa{ \Phi^a \rightarrow v \hat r^a - {\hat r^a \over er } . }
This power law falloff is due to the massless dilaton field in this
scale invariant limit.  To define the dilaton field ${\cal D}$
we write fluctuations
of $\Phi^a $ about the asymptotic monopole configuration in the form
\eqn\twofb{\Phi^a = v \hat r^a e^{\cal D} = v \hat r^a + v \hat r^a
{\cal D } + \cdots}
We can then define a dimensionless ``dilaton charge'' as
\eqn\twofc{ Q_{dil} = v \int_{S^3_\infty} \vec \nabla {\cal D}
 \cdot d \vec S }
and using \twofa\ we see that for the monopole solution
$Q_{dil} = 4 \pi /e = g = M_M/v $.

\subsec{Collective Coordinates and the Monopole Moduli Space}

Given a classical solution in field theory one often finds that the solution is
part of a multi-parameter family of solutions with the same energy.  The
parameters
labeling the different degenerate solutions are called collective coordinates
or moduli and the space of solutions of fixed energy (and topological charge)
is called the moduli space of solutions.  Before discussing the general
situation it will be useful to identify the four collective coordinates of a
charge
one BPS monopole \osbornb.

To start with, in \twof\ we have constructed a monopole sitting at the origin.
By translation invariance of \onen\ a monopole sitting at any other point in
$R^3$ is also a solution with the same energy. If we let $\vec X$ denote this
center of mass collective coordinate then the general solution is
\eqn\twog{ \Phi^a_{cl} (\vec r + \vec X), \qquad {A^a_i}_{cl} (\vec r + \vec X)
}
with the classical solutions at $\vec X=0$ given by \blatz. We can construct
a slowly moving monopole by letting $X$ depend on time so that we have
fields ${A^a_i}_{cl} (\vec r + \vec X (t))$. This time dependence will
of course give rise to an electric field and the energy of the monopole
will exceed the Bogomol'nyi bound by the kinetic energy of the
monopole.  This is precisely what should happen for motion in the
moduli space, the potential terms stay constant and the kinetic
terms are proportional to the velocity of the motion along the
moduli space.

The remaining collective coordinate is somewhat more subtle.  At this
point it is useful to recall some basic facts about the configuration space of
gauge theories.  In gauge theory it is important to make a distinction between
{\it small} gauge transformations $g$ which are those approaching the
identity at spatial infinity and {\it large} gauge transformations which do not
approach the identity at spatial infinity. These play different roles in gauge
theory. In particular given the space of gauge and Higgs fields ${\cal A} =
(A,\Phi)$
the physical configuration space is given by
\eqn\twoh{ {\cal C} = {\cal A}/ {\cal G} }
where ${\cal G}$ is the group of small gauge transformations. Thus physical
states
are invariant under small gauge transformations and they do not act as
symmetries
of ${\cal C}$.  Rather, they describe a redundancy in our description of the
theory
when we work just in ${\cal A}$. Large gauge transformations on the other hand
do not identify points in ${\cal C}$
but instead act as true symmetries which relate different points in ${\cal C}$
with
the same properties.

With this in mind we will try to identify an additional collective coordinate
associated to global $U(1)$ electromagnetic gauge transformations.
Heuristically we expect such a collective coordinate because the monopole
solution contains excitations of the electrically charge $W^\pm$ fields in
its core.

We will work in $A_0=0$ gauge with a BPS monopole configuration $A_i, \Phi$
obeying the equation $B_i = D_i \Phi$. A deformation of this solution $\delta
A_i(\vec x,t),
\delta \Phi(\vec x,t)$ which keeps
the potential energy fixed must obey the linearized Bogomol'nyi equation
\eqn\twoi{\epsilon_{ijk} D_j \delta A_k = D_i \delta \Phi + [ \delta A_i \phi]
}
and the Gauss law constraint
\eqn\twoj{D_i \delta {\dot A}_i + [ \Phi, \delta {\dot \Phi} ] = 0 }
The unique solution (modulo small gauge transformations) is
\eqn\twok{\eqalign{ \delta A_i & = D_i (\chi (t) \Phi) \cr
                                     \delta \Phi  & = 0 \cr
                                     \delta A_0 & = D_0 (\chi (t) \Phi ) - \dot
\chi \Phi \cr }}
where $\chi(t)$ is an arbitrary function of time. Note that $A_0 $ vanishes
identically,
it has been written in the form \twok\ to make clear the relation with
gauge transformations. This solution has the following
properties

\item{1.} It obeys \twoi and \twoj
\item{2.} For $\dot \chi = 0$   the deformation is by a large gauge
transformation with
$g = e^{\chi \Phi}$ so that $\chi$ is a physical zero mode.
\item{3.} For $\dot \chi \ne 0$ the linearized Bogomol'nyi equation is still
satisfied so there
is no change in the potential energy ($\vec B^2 + (\vec D \Phi)^2 $) but there
is
an increase in the kinetic energy ($ \vec E^2  $).  If we think of the
configuration space
as a mountain range then the moduli space is a flat valley.  Motions that stay
purely along
the valley have fixed potential energy but variable kinetic energy, as we have
found.
\item{4.} Since the unbroken gauge group $U(1)$ is compact, $\chi$ is
a periodic coordinate. Therefore the one monopole moduli space is
topologically ${\cal M}_1 = R^3 \times S^1$.

So far we have limited our discussion to monopoles with $N=1$. It is at first
sight not clear whether we  expect static solutions to exist with $N \ge1 $ and
if they
do what the collective coordinates should be. Physically, we can argue as
follows.
Away from the BPS limit the photon is the only massless field in the theory.
Multi-monopoles
of the same sign magnetic charge which are well separated
will thus experience a
Coulomb repulsion and we thus do not expect static solutions for such a
configuration.
On the other hand in the BPS limit the Higgs field is really a dilaton of
spontaneously
broken scale invariance (at least classically) and we have seen that the one
monopole
solution carries a charge under this Higgs field.  Since Higgs exchange is
always attractive,  there can be a cancellation between the Coulomb repulsion
and Higgs attraction. Classically this cancellation does occur as a
consequence of the fact that the magnetic charge and dilaton charge
of the monopole are equal as was found below \twofc.  This equality
is not accidental but has its roots in spontaneously broken scale
invariance  which forces the dilaton charge of any state to equal its
mass.

Thus we might expect on physical grounds that there are solutions given by
well separated static monopoles and that for magnetic charge $k$ the moduli
space is $4k$ dimensional with the collective coordinates being the locations
of the $k$ monopoles and their dyon degrees of freedom. This is correct, at
least
for large separation, although the above hardly constitutes a serious argument.
A careful analysis of the issue would take us to far afield. It basically
involves
the use of index theory to count perturbations of the Bogomol'nyi equations.
A discussion suitable for physicists may be found in \ewein, a mathematical
proof is given in \taubes.

Suitably explicit  multimonopole solutions are hard to come by but in spite of
this it is possible to say some general things about the structure of the
multi-monopole
moduli space. In the final lecture we will discuss the structure of the
two-monopole
moduli space in some detail.

Before proceeding it is useful to make use of a connection between the
Bogomol'nyi equations on $R^3$ and the self-dual Yang-Mills equations
on $R^4$.  If we write $A_4 = \Phi$ then we can rewrite the Bogomol'nyi
equation
\twod\ as
\eqn\twol{ F_{ab} = *F_{ab} }
where $a,b=1 \cdots 4$, we work on $R^4$ with coordinates $x_1, x_2, x_3, x_4$
and Euclidean signature so that $**=1$, and we restrict ourselves to
configurations
which are independent of $x_4$.   This suggest that there is a deep connection
between the problem of solving the Bogomol'nyi equations and the problem
of solving the self-dual  Yang-Mills equations.
We will just use this
connection to simplify the notation.  For example Gauss' Law reads
\eqn\twom{ D_a {\dot A_a} = 0}
and gauge transformations take the form
\eqn\twon{ \delta A_a = D_a \Lambda }
where it is always understood that all quantities are independent of $x_4$.

Now the $k$ monopole moduli space $ {\cal M}_k$ is defined as the space of
solutions
to the Bogomol'nyi equations having topological charge $k$. Tangent vectors
to ${\cal M}_k$, $\delta_\alpha A_a $, are deformations of a given charge $k$
solution $A_a \rto A_a + \delta_\alpha A_a $ which satisfy the linearized
Bogomol'nyi equations
\eqn\twoo{ D_a \delta_\alpha A_b - D_b \delta_\alpha A_a = \half
\epsilon_{abcd} ( D_c \delta_\alpha A_d - D_c \delta_\alpha A_c ) }
and are orthogonal to (small) gauge transformations
\eqn\twop{ D_a \delta_\alpha A_a = 0 }
so that they leave one
in the physical configuration space.

Given such a tangent vector the metric on ${\cal M}_k$ is
\eqn\twoq{ {\cal G}_{\alpha \beta} = - \int d^3 x {\rm Tr} \delta_\alpha A_a
\delta_\beta A_a }

This metric is inherited from the action for the underlying gauge theory.
To see this, imagine we are given a charge $k$ BPS monopole solution
$A_a(\vec x, z^\alpha )$ depending on $4k$ collective coordinates
$z^\alpha$.  By definition the potential energy is independent of the
$z^\alpha$.  Now one might think that we could construct tangent vectors
to ${\cal M}_k$ simply by differentiating with respect to the $z^\alpha$.
This is not quite correct because there is no guarantee that the resulting
change to $A_a$ is orthogonal to gauge transformations, in other words
differentiating with respect to the $z^\alpha$ may include a gauge
transformation. However we can always cure this by undoing the gauge part of
this variation by writing the tangent vector as
\eqn\twor{\delta_\alpha A_a = {\partial A_a \over \partial z^\alpha} - D_a
\epsilon_\alpha }
where $\epsilon_\alpha(\vec x,z^\beta)$ is a gauge parameter chosen to ensure
that \twop\ is satisfied.

To construct the metric we consider slow time dependent
variations of the collective coordinates $z^\alpha(t)$.  If we
write
\eqn\twos{ A_a(\vec x, z^\alpha (t) , \quad A_0 = {\dot z}^\alpha
\epsilon_\alpha }
then $F_{0 a} = {\dot z}^\alpha \delta_\alpha A_a $
and the action is
\eqn\twot{ S = - {1 \over 2} \int d^3 x d t {\rm Tr}  F_{0a} F^{0a}  = {1 \over
2}
\int dt {\cal G}_{\alpha \beta} {\dot z}^\alpha {\dot z}^\beta }
with ${\cal G}_{\alpha \beta}$ as in \twoq.

\subsec{Exercises for Lecture 2}
\item{E4.} For a dyon with electric and magnetic charge $(q,g)$ prove the
bound
\eqn\exfour{ M \ge v (q^2 + g^2 )^{1/2}. }
\item{E5.} Following the discussion in the lecture derive the action for the
collective coordinates $\vec X, \chi$ of the charge one BPS monopole.
Quantize this action to deduce the spectrum of states in the magnetic
charge one sector. Note that the states you obtain
by quantizing the dyon collective coordinate $\chi$ consist of an
infinite tower of states of increasing mass and electric charge.
Thus in the monopole sector electric charge is classically continous,
but is quantized when treated quantum mechanically.
\item{E6} Compute the generators $J^i$ and $T^a$ of rotations and global
gauge transformations and verify that the ansatz \blatz\ is left
invariant by the action of $\vec J + \vec T$.
\item{E7.} It is possible to view the gauge parameter $\epsilon_\alpha (x,z)$
as
a connection on ${\cal M}_k$ with covariant derivative $s_\alpha =
\partial_\alpha
+ [\epsilon_\alpha, \quad ] $. Show that  $\delta_\alpha A_a$ can then be
viewed as a mixed component of the curvature of the connection $(A_a,
\epsilon_\alpha)$
on $R^4 \times {\cal M}_k $.
\item{E8.} Compute the dilaton charge of a massive $W^+$ boson at
rest at the origin and show that it is equal to $m_W/v$. Show that
this follows from the general theory of spontaneously broken scale
invariance.

\newsec{Lecture 3}

\subsec{Witten Effect}
There is a famous term, the  $\theta$ term, which can be added to the
Lagrangian
for Yang-Mills theory without spoiling renormalizability. It is given by
\eqn\threea{ {\cal L}_\theta = - {\theta e^2 \over 32 \pi^2} F_{\mu \nu}^a
*F^{a \mu \nu } .}
This interaction violates $P$ and $CP$ but not $C$.  Since it preserves $C$ we
may expect that it is consistent with the existence of a duality symmetry of
the theory.
As is well known \coleref,  this term is a surface term and does not affect the
classical equations of motion.  There is however $\theta$ dependence in
instanton
effects which involve non-trivial long-range behavior of the gauge fields. As
was
realized by Witten \witteff,  in the presence of magnetic monopoles $\theta$
also has
a non-trivial effect, it shifts the allowed values of electric charge in the
monopole
sector of the theory.

I will give two explanations of this effect, the first is borrowed from Coleman
\coleb,
the second from Witten \witteff.
First consider pure electromagnetism. Then the $\theta$ term reduces to the
QED interaction
\eqn\threeb{ {\cal L}_\theta = {\theta e^2 \over 8 \pi^2} \vec E \cdot \vec B .
}
Now consider this interaction in the presence of a (Dirac) magnetic monopole.
Writing the fields as a monopole field plus corrections we have
\eqn\threec{\eqalign{\vec E & = \vec \nabla A_0 \cr \vec B & = \vec \nabla
\times \vec A
+ {g \over 4 \pi} {\hat r \over r^2 } . \cr }}
Substituting into the action density \threeb\ we obtain
\eqn\threed{\eqalign{ L_\theta =&  \int d^3 r {\cal L}_\theta = {\theta e^2
\over 8 \pi^2}
\int d^3 r \vec \nabla A_0 \cdot (\vec \nabla \times \vec A + {g \over 4 \pi}
{\hat r \over r^2} ) \cr
 = & - {\theta e^2 g \over 32 \pi^3} \int d^3 r A_0 \vec \nabla \cdot {\hat r
\over r^2} =
- {\theta e^2 g \over 8 \pi^2} \int d^3 r A_0 \delta^3 (\vec r) } }
which we recognize as the coupling of the scalar potential $A_0$ to an electric
charge
of magnitude $-\theta e^2 g / 8 \pi^2 $ located at the origin. In other words,
the
magnetic monopole has acquired an electric charge. For a minimal charge
monopole with $eg=4 \pi$ the electric charge of the monopole is $-e \theta / 2
\pi$. Although this derivation gives the correct answer, one may feel
a bit uneasy about the method used. We don't really know what is going
on at the origin for a Dirac monopole yet this calculation suggests
a delta function electric charge density located at the origin.

A more fundamental derivation which applies to the full
$SU(2)$ gauge theory and  which does not suffer from this ambiguity runs as
follows.
We have seen that the dyon collective coordinate of the monopole allows
it to carry electric charge. The dyon collective coordinate arises through
$U(1)$ gauge
transformations which are constant at infinity.  We now consider these
transformations
in the presence of a theta term. We are interested in gauge transformations,
constant
at infinity, which are rotations in the $U(1)$ subgroup of $SU(2)$ picked out
by the
gauge field. That is, rotations in $SU(2)$ about the axis $ {\hat \Phi}^a =
\Phi^a/|\Phi^a|$.
The action of such  an infinitesimal gauge transformation on the field is
\eqn\threee{\delta A_\mu^a = {1 \over ev} (D_\mu \Phi)^a }
with $\Phi$ the background monopole Higgs field.  Let ${\cal N}$ denote the
generator of this gauge transformation. Then if we rotate by $2 \pi$ about
the $\hat \Phi$ axis we must get the identity \foot{At this point we are
working in
a theory with gauge group $SU(2)/Z_2 = SO(3)$ since all states are in the
adjoint
representation. In this theory a $2 \pi$ rotation gives the identity.  Later,
when
we consider $SU(2)$ and states in the fundamental representation this condition
will be modified since then a $2 \pi$ rotation gives an element
of the center of $SU(2)$ which acts non-trivially on the fundamental
representation.}.
That is, physical states must obey
\eqn\threef{e^{2 \pi i {\cal N}} = 1. }

It is straightforward to compute ${\cal N}$ using the Noether method,
\eqn\threeg{{\cal N} = { \partial {\cal L}  \over \partial \partial_0 A_\mu^a }
\delta A_\mu^a }
with $\delta A_\mu^a$ given by \threee. Including the theta term one finds
\eqn\threeh{ {\cal N} = {Q \over e} + {\theta e g \over 8 \pi^2} }
where
\eqn\threei{\eqalign{ g & = {1 \over v} \int d^3 x D_i \Phi^a B_i^a \cr
                                       Q & = {1 \over v} \int d^3 x D_i \Phi^a
E_i^a \cr }}
are the magnetic and electric charge operators respectively.  The condition
\threef\
thus implies that
\eqn\threej{ Q = n_e e - {e \theta n_m \over 2 \pi} }
where $n_e$ is an arbitrary integer and $n_m = eg/4 \pi$ determines the
magnetic charge
of the monopole.

\subsec{Montonen-Olive and SL(2,Z) Duality}
Let us pause for a moment to see what we have accomplished in trying to
establish a duality between electric and magnetic
degrees of freedom. In the BPS
limit  at $\theta = 0$ we have a classical spectrum indicated in the table
below

\centerline{Table 1}
$$\vbox{ \settabs 4 \columns
\+ & Mass & $(Q_e,Q_m)$ & Spin \cr
\+ Higgs & ~~ 0 & ~~ $(0,0)$ & ~~ 0 \cr
\+ Photon & ~~ 0 & ~~ $(0,0)$ & ~~ 1 \cr
\+ $W^\pm $ & ~~ ve & ~~ $(e,0)$ & ~~ 1 \cr
\+$ M^{\pm} $ & ~~ vg & ~~ $(0,g)$ & ~~ 0 \cr } $$

As is evident from the table, all of these states saturate the
Bogomol'nyi bound
$M \ge v \sqrt{ Q_e^2 + Q_m^2} $ with $Q_m= 4 \pi n_m/e$ and
$Q_e = n_e e - e n_m \theta / 2 \pi$ (with $\theta = 0$ for the moment).
At weak coupling, where this analysis should be a good first order
approximation to the full quantum answers, we have
\eqn\threek{ M_W = e v << v, \quad M_M = gv = {4 \pi \over e} v >> v }
so although we have constructed a theory with both electric and magnetic
charges, monopoles are much heavier than $W$ bosons at weak coupling.
However we would expect that to get a dual theory we would also have to
exchange the role of electric and magnetic charge. Given the quantization
condition
this implies that we should look for a duality transformation which acts on the
fields as in \oneb\ but also takes
\eqn\threel{ e \rto g \equiv { 4 \pi \over e} }
and relabels electric and magnetic states.

Based on the classical spectrum shown in Table 1 and some other arguments
Montonen and Olive proposed that this should be an exact duality of the $SO(3)$
Yang-Mills-Higgs theory in the BPS limit \MO.  However, as noted by the authors
of \MO, there are some obvious problems with this proposal. They are:
\item{1.} Quantum corrections would be expected to generate a non-zero
potential
$V(\Phi)$ even if one is absent classically and should also modify the
classical mass formula.
Thus there is no reason to think that the duality of the spectrum should be
maintained
by quantum corrections.
\item{2.} The $W$ bosons have spin one while the monopoles are rotationally
invariant indicating that they have spin zero. Thus even if the mass spectrum
is
invariant under duality, there will not be an exact matching of states and
quantum numbers.
\item{3.} The proposed duality symmetry seems impossible to test since rather
than acting
as a symmetry of a single theory it relates two different theories, one of
which is necessarily
at strong coupling where we have little control of the theory.

As we will see later, the first two problems are resolved by embedding the
theory
into $N=4$ super Yang-Mills theory \osborn.
The third problem is still with us in that there
are few concrete ways to test the proposal. However there
{\it are} non-trivial tests
and the first of these arose by first considering an extension of duality to a
larger
set of transformations.

It is not hard to see that if the basic duality idea is correct then it should
have an
interesting  extension when the effects of a non-zero theta angle are included.
Including the theta term, the Lagrangian we are considering is determined by
two real parameters, $e$ and $\theta$.  We can write the Lagrangian as
\eqn\threem{ \eqalign{ {\cal L} = & - {1 \over 4} F^{\mu \nu} F_{\mu \nu}
 - {\theta e^2 \over 32 \pi^2} F^{\mu \nu} *F_{\mu \nu} -
{1 \over 2} D^\mu \Phi D_\mu  \Phi \cr & \equiv - {1 \over 32 \pi}
\Im ( {\theta \over 2 \pi} +
{4 \pi i \over e^2} ) (F^{\mu \nu} + i *F^{\mu \nu})(F_{\mu \nu} +
i *F_{\mu \nu} )  - {1 \over 2} D^\mu  \Phi D_\mu  \Phi \cr }}
We thus see that the Lagrangian can
be written in terms of a single complex  parameter
\eqn\threen{ \tau = {\theta \over 2 \pi} + {4 \pi i \over e^2 }. }
As an aside, note that $n$ instanton effects in this theory are weighted by
$e^{2 \pi i n \tau}$.

Since physics is periodic in $\theta$ with period $2 \pi$  the
transformation
\eqn\threeo{ \tau \rto \tau + 1}
should leave physics invariant up to a relabeling of states. At $\theta=0$ the
duality transformation \threel\  is given in terms of $\tau$ by
\eqn\threep{ \tau \rto - {1 \over \tau} .}
It thus seems reasonable to suspect that at arbitrary $\theta$ the full duality
group
is generated by transformations of the form \threeo\ and \threep.   It is a
well known
fact that these two transformations generate the group $SL(2,Z)$ of projective
transformations
\eqn\threeq{ \tau \rto {a \tau + b \over c \tau + d }, \quad a,b,c,d \in Z,
\quad ad-bc=1. }

Note that since $e^2 \ge 0$,  $\tau$ naturally lives on the upper half plane
$\Im \tau \ge 0$. Furthermore, one can check that one can use $SL(2,Z)$
transformations to  map any $\tau$
in the upper half plane into the fundamental region defined by
$-1/2 \le {\rm Re} \tau \le 1/2$ and $ |\tau| > 1$.

In order for \threeq\ to be a symmetry we know that there must in addition be
a relabeling of states. From $\threej$ we see that the transformation  \threeo\
shifts
the electric charge by $-1$ (for $n_m-1$)
 and we know from the earlier discussion that the transformation
\threep\  requires an exchange of electric and magnetic quantum numbers.
Putting
these two facts together we deduce that the action of  $SL(2,Z)$ on the quantum
numbers should be
\eqn\threeqa{ \left (\matrix{
n_e\cr n_m\cr }\right ) \rto \pmatrix{ a & -b \cr c & -d \cr }\left (\matrix{
n_e\cr n_m\cr }\right ) }

Finally, let us consider the spectrum of states saturating the BPS bound
$M^2 \ge v^2 (Q_e^2 + Q_m^2)$. We know the allowed values of $Q_e$
and $Q_m$ are
\eqn\threer{\eqalign{ Q_m &  = {4 \pi \over e} n_m \cr
  Q_e & = n_e e - n_m { e \theta \over 2 \pi } . \cr }}
Substituting these into the formula for $M^2$ and writing the result in terms
of
$\tau$ yields
\eqn\threes{ M^2 \ge 4 \pi v^2 \left (\matrix{
n_e, & n_m \cr }\right ) {1 \over \Im \tau}  \pmatrix{
1 & - \Re \tau \cr
- \Re \tau & | \tau |^2 \cr } \left (\matrix{ n_e\cr n_m\cr }\right ) }
It is left as an exercise to verify that the mass formula in this form is
invariant
under $SL(2,Z)$ transformations.

The extension of electromagnetic duality to $SL(2,Z)$ that we have uncovered is
usually referred to as $S$-duality. The name is a historical accident. Although
this extension was first discovered in the context of lattice models \cardy\ it
was
discussed as a symmetry of $N=4$ Yang-Mills theory first  in the low-energy
limit of toroidal compactifications
of string theory \sdual.  In that context the variable $\tau$ becomes a
dynamical field
usually denoted by $S$ and the $SL(2,Z)$ transformations of $S$ were called
$S$-duality to distinguish it from other $SL(2,Z)$ transformations in string
theory
which are (superficially) unrelated.

\subsec{Exercises for Lecture 3}
\item{E9.}Carry out the computation of the generator
${\cal N}$ using the Noether method.
Verify that dyons with electric charge
$Q = n e - e \theta / 2 \pi $ satisfy
the DSZ quantization condition but violate $CP$.
Show that the interaction ${\cal L}_\theta$
violates $CP$.
\item{E10.} Find two points on the boundary of the fundamental region
described in the text
which are left fixed by some element of $SL(2,Z)$ other
than the identity. What order are these elements of
$SL(2,Z)$ and what are the values of $\theta$ and $e^2$ at the two fixed
points?
\item{E11.} Show that $M^2$ is left invariant by the $SL(2,Z)$ transformation
given
by \threeq\ and \threeqa. For those familiar with string theory note
the close connection between the form of $M^2$ and the Poincare metric
on the upper half plane.
\item{E12.} I have been sloppy about the precise group which is acting
in \threeq\ and \threeqa. Show that there are order two elements which
act trivially on the couplings but non-trivially on the charges as
in \threeqa.  What do these order two elements correspond to
physically?

\newsec{Lecture 4}

\subsec{Monopoles and fermions}
As we have seen in the previous discussion, duality is inherently a quantum
symmetry since it relates weak coupling to strong coupling.
As such we cannot hope to understand it easily unless we are working in a
theory
where quantum effects are under rather precise control. At our current level of
understanding this limits us to theories with supersymmetry, and the more
supersymmetry,
the more control we have of the dynamics. Supersymmetry involves the addition
of fermion fields with special couplings.  However many of the features of
fermions
in monopole backgrounds are independent of supersymmetry. Thus we will start
out
with a general discussion of the effects of fermions and then later generalize
our results to the supersymmetric context.

We will first consider Dirac fermions with couplings
to the fields appearing in \onen\ determined by the
Lagrangian
\eqn\foura{ {\cal L}_\psi =
i {\bar \psi}_n \gamma^\mu (D_\mu \psi)_n - i {\bar \psi}_n
T^a_{nm} \Phi^a \psi_m }
with $T^a_{nm}$ the anti-Hermitian generators of $SU(2)$ in the representation
${\bf r}$.
We will consider only fundamental and adjoint fermions in which case we take
\eqn\fourb{ T^a_{nm} = - {i \over 2} \tau^a_{nm} \quad n,m=1,2 }
with $\tau^a$ the Pauli matrices
or
\eqn\fourc{ T^a_{nm} = \epsilon^a_{nm} \quad n,m=1,2,3 .}
It will also be convenient following \jackiwr\
to use a representation of the gamma matrices with
\eqn\fourd{ \gamma^0 =  \pmatrix{
0 & -i \cr
i & 0 \cr
} \qquad \gamma^i =  \pmatrix{
-i \sigma^i & 0 \cr
0 & i \sigma^i  \cr
} }
obeying $\{ \gamma^\mu, \gamma^\nu \} = 2 \eta^{\mu \nu}$.  We will also
 write $\alpha^i = \gamma^0 \gamma^i $.

The Dirac equation is then
\eqn\foure{ ( i \gamma^\mu D_\mu - \Phi ) \psi = 0 . }
For a monopole configuration with $A_0=0$ we can look for stationary solutions
of the form $\psi(\vec x,t) = e^{i E t} \psi (\vec x )$. Writing $\psi$ in
terms of
two-component spinors
\eqn\fourf{ \psi = \left (\matrix{ \chi^+\cr \chi^- \cr } \right ) }
we then have the coupled equations
\eqn\fourg{\eqalign{  \Dsl  \chi^- & \equiv
        (i \sigma^i D_i +  \Phi ) \chi^-   = E \chi^+  \cr
            {\Dsl}^\dagger \chi^+ &
        \equiv ( i \sigma^i D_i -  \Phi ) \chi^+ = E \chi^-  \cr }}

Now \fourg\ will have solutions with $|E| >0 $ and may also have solutions
with $E=0$. We can quantize the fermion fluctuations about a magnetic monopole
by expanding $\psi$ in terms of eigenfunctions of the Dirac operator \fourg\
and then interpreting the coefficients multiplying the eigenfunctions as
creation
and annihilation operators with anti-commutation relations which follow from
the canonical anti-commutation relations of $\psi$. Modes with $|E|>0$ will
thereby
lead to configurations with energy greater than the ground state energy of the
monopole.
On the other hand, if \fourg\ has solutions with $E=0$ then the states created
by
the corresponding creation operators will be degenerate in energy with
the original monopole solution. Thus we can view the zero energy eigenfunctions
of the Dirac operator as ``fermionic collective coordinates'' in the sense that
they describe Grassmann valued deformations of the monopole which keep the
energy fixed.

Thus to study the structure of the monopole ground state we must study the zero
energy
solutions of \fourg, that is we want the solutions of $\Dsl \chi^+=0$ and
$\Dsl^\dagger \chi^- =0$,
the kernels of $\Dsl$ and $\Dsl^\dagger$.  It is easy to see that the ${\rm
ker}{ \Dsl}^\dagger = \{ 0 \} $
using the fact that $ {\rm ker} {\Dsl}^\dagger \subset {\rm ker} \Dsl
{\Dsl}^\dagger $ and that
$ \Dsl {\Dsl}^\dagger $ is a positive definite operator.  On the other
hand ${\rm ker} \Dsl$ is non-zero in a monopole background and can be computed
using an index theorem of Callias \call\ which gives
\eqn\fourh{ {\rm dim ~ ker} \Dsl - {\rm
dim ~ ker } {\Dsl}^\dagger = A(r) n_m }
 with $n_m$ the winding number of the Higgs field
(the monopole charge) and $A(r)$  a constant depending on the representation
of the fermion fields and the ratio of the magnitude of a bare fermion mass to
the
Higgs expectation value. In the examples we are discussing $A=1$ for
fundamental
fermions and $A=2$ for adjoint fermions.

While both fundamental and adjoint fermions have zero modes in a monopole
background, their consequences are somewhat different so we discuss the
two cases separately.

\subsec{Monopoles coupled to isospinor fermions}

For fundamental fermions in the ${\bf 2}$ of $SU(2)$ a charge
one monopole has a
single fermion zero mode according to \fourh.  To be precise, there is a single
zero mode wave function but since the fermion $\psi$ does not obey any reality
condition, the coefficient multiplying the zero mode should be taken complex.
We thus have the expansion
\eqn\fouri{ \psi = a_0 \psi_0 + \hbox{non-zero modes} }
and the anti-commutation relations for $\psi$ imply
\eqn\fourj{\{a_0^\dagger,a_0 \} = 1, \quad \{a_0, a_0 \} = \{ a_0^\dagger,
a_0^\dagger \} = 0 . }
To construct the monopole ground state we start with a ground state $ | \Omega
\rangle$
with $a_0 | \Omega \rangle = 0$ and then act with $a_0^\dagger$.  This gives a
two-fold
degenerate ground state consisting of the two states.
\eqn\fourk{  | \Omega \rangle, \qquad
        a_0^\dagger  | \Omega \rangle }

Given this degeneracy it is natural to ask whether there are quantum numbers
which
distinguish the two ground states.  The standard answer, given in
\jackiwr, is that this theory has a fermion number conjugation
symmetry
\eqn\fourka{ \psi_n \rightarrow \pmatrix{ \sigma^2 & 0 \cr 0 & -\sigma^2 \cr }
\tau_{nm}^2  \psi_m^* }
which changes the sign of the $U(1)$ fermion number charge of any state.
The two degenerate monopole states differ by one unit of fermion number
since one obtains one from the other by acting with fermion
creation operators which carry fermion number one. On the other hand
the fermion number conjugation symmetry can only be respected if the
two states carry opposite fermion number.
Thus it is argued in \jackiwr\ that the two states in \fourk\ have
fermion number $\pm 1/2$. However this argument suffers from the following
difficulty. The discrete symmetry \fourka\ is a classical symmetry of
the theory which forbids a Dirac mass term for the matter fermions.
On the other hand one can show that instanton effects will generate
such a Dirac mass term which is related to the fact that the
symmetry \fourka\ involves a discrete chiral transformation. In
other words, the fermion number conjugation symmetry is anomalous and is not
an exact quantum symmetry of the full theory. However, at $\theta=0$
the theory is also $CP$ invariant. $CP$ takes the magnetic charge
into itself but changes the sign of the fermion number charge. Thus
$CP$ invariance enforces the assignment of fermion number $\pm 1/2$
to the two ground states.

A more interesting example of charge fractionalization occurs if we take
$N_f$ flavors of Dirac fermion coupled to a monopole background with Lagrangian
\eqn\fourl{ {\cal L}_\psi = \sum_{I=1}^{N_f} i {\bar \psi}^I \gamma^\mu D_\mu
\psi^I
- i f {\bar \psi }^I \Phi \psi^I }
In this theory the $U(1) \sim O(2)$ fermion number symmetry is extended to a
$O(2N_f)$
symmetry.  Of this symmetry only a $SU(N_f) \times U(1) $ symmetry is manifest
in
the Lagrangian \fourl. To see the full $O(2 N_f)$ symmetry note that we can
write
the Dirac fermions $\psi^I$ in terms of $ 2 N_F$ Weyl fermions $\chi^a , a=1
\ldots 2 N_f$.
Furthermore, since the doublet representation of $SU(2)$ is pseudoreal we can
take the $\chi^a$ to all be say left-handed under the Lorentz group and to all
transform
the same way under the $SU(2)$ gauge symmetry.  If we write \fourl\ in this
basis
then the quadratic  terms involving the $\chi^a$ are $O(2 N_f)$ invariant and
$O(2 N_f)$ commutes with both the Lorentz group and the gauge group.

Thus we have $2 N_f$ Weyl fermions transforming as a vector of $O(2 N_f)$ and
in
the zero mode expansion of $\psi^I$ we will have
\eqn\fourm{ \psi^I = a_0^I \psi_0 + \hbox{non-zero modes} }
To see what the consequences are for the spectrum it is useful to first
rephrase the results we found for $N_f=1$. There we could trade the
operators $(a_0,a_0^\dagger)$ for a pair of self-conjugate operators
$(b_0^1, b_0^2)$ by writing
\eqn\fourn{\eqalign{ a_0 & = {1 \over \sqrt{2}} (b_0^1 + i b_0^2 ) \cr
                                      a_0^\dagger & = {1 \over \sqrt{2}}
(b_0^1 - i b_0^2 ) \cr }}
The $b_0^a$, $i=1,2$ then obey the Clifford algebra
\eqn\fouro{ \{ b_0^i, b_0^j \} = \delta^{ij}. }
The ground state must furnish a representation of this Clifford algebra
and since the smallest representation is two-dimensional we again conclude
that the ground state is two-fold degenerate.

We can apply the same technique for arbitrary $N_f$ in which case we end
up with operators $b_0^a$, $a = 1 \cdots 2 N_f$ obeying
\eqn\fourp{ \{ b_0^a, b_0^b \} = \delta^{ab}. }
Representations of this Clifford algebra have dimension $2^{2 N_f/2} =
2^{N_f}$,
that is the monopole ground state is now a spinor of $SO(2 N_f)$!
This is precisely the phenomenon that allows one to construct spacetime
fermions in the Ramond sector of superstring theory.

Since we have added fermions and changed the global structure of
the gauge group (from $SO(3)$ to $SU(2)$ ) we should also go back
and reanalyze the constraint \threef\ which followed from the action of
global $U(1)$ charge rotations.  We can repeat most of the previous discussion
but with one change, since we are now working in $SU(2)$ and not $SO(3)$
a rotation by $2 \pi$ about some axis does not give the identity but rather
gives the non-trivial element of the center of $SU(2)$ which acts on spinor
representations as $-1$. If we denote this operator by $(-1)^H$ following
\swtwo\ then we have the relation
\eqn\fourq{ \exp ( 2 \pi i ( {Q \over e} + {\theta n_m \over 2 \pi} ) ) =
(-1)^H }
If $Q = n_e e - e n_m \theta / 2 \pi$ then \fourq\ says that there is a
correlation
between the action of the center of $SU(2)$ and the electric charge,
states in spinor representations of $SU(2)$ have $n_e$ half an odd integer
and states transforming trivially under the center must have $n_e$ integer.
This of course agrees with our expectations in the zero magnetic charge
sector of the theory.  In the monopole sector the implications of \fourq\ are
as follows.  Since $(-1)^H$ acts as the center of $SU(2)$ and the fermion
fields $\psi^I$ are doublets of $SU(2)$, $(-1)^H$ acts to change the sign of
the fermion fields. That is
\eqn\fourr{ (-1)^H \psi^I (-1)^H = - \psi^I }
or $\{ (-1)^H, \psi^I \} = 0$.
In the monopole sector, after expanding in zero modes we will then have
$\{ (-1)^H, b_0^a \} = 0 $ and we must represent the action of $(-1)^H$ on the
$2^{N_f}$ fold degenerate spectrum and impose the constraint \fourq.
But this is a completely familiar problem. We can think of the $b_0^a$ as
gamma matrices and $(-1)^H$ as the analog of ``$\gamma^5$ ''.  In other
words, the spinor representation of $SO(2N_f)$ of dimension $2^{N_f}$
is reducible and splits into two irreducible representations, each of dimension
$2^{N_f -1}$ with eigenvalues $\pm 1$ under $(-1)^H$. Thus we learn
from \fourq\ that in the monopole sector there is a correlation between
the electric charge of dyon states and their transformation properties
under the global $SO(2 N_f)$ symmetry. As discussed in \swtwo\ this
is also required physically in order that one not make states
in monopole - antimonopole annihilation which do not occur in the
perturbative spectrum.

One particularly interesting example of this phenomenon occurs for
$N_f = 4$.  Then by the above analysis,  the fermion fields $\psi^I$
carry electric charge $e/2$ and are in the eight-dimensional vector
representation,
$8_v$, of the global $SO(8)$ symmetry (which we should really call $Spin(8)$
since
there are spinors in the monopole sector.). On the other hand in the one
monopole sector  the spinor of dimension $2^4=16$ splits into two
eight-dimensional spinor representations $16 \rto 8_s + 8_c$ and
from the constraint \fourq\ we see that the neutral monopole (or evenly
charged dyons)  transforms
as $8_s$ while the odd charged dyons transform as $8_c$. Thus there
seems to be a $Spin(8)$ triality as well as a possible electromagnetic duality
in this theory, at least classically. In fact, when embedded into $N=2$
supersymmetric
gauge theory, this theory does appear to be self-dual with the $SL(2,Z)$
duality
group extended to a triality action on $Spin(8)$ \refs{\swtwo, \gaunharv,
\sethi}.

\subsec{Monopoles coupled to isovector fermions}

If we take the fermions in the adjoint representation then the index theorem
\fourh\ predicts two zero modes for a charge one monopole.  Besides a doubling
of the number of zero modes there is one other important difference from the
isospinor case which involves the spin carried by the fermion zero modes.
This can be understood as follows. From the discussion in sec. 2.1 we saw
that the angular momentum generator for a symmetric monopole is
\eqn\fours{ \vec K = \vec L + \vec S + \vec T }
with $\vec L + \vec S$ the sum of orbital and spin terms generating the usual
rotation group and $\vec T$ the  $SU(2)$ generators. That is, the $SU(2)$
invariance
group is a diagonal subgroup of the usual rotation group $SU(2)_R$ and the
gauge group $SU(2)_G$.

Now isospinor fermions in the ${\bf 2}$ of $SU(2)_G$ can have $K=0$ since
${ \bf 2} \times {\bf 2} = {\bf 3} + {\bf 1} $ and that is consistent with the
fact that the zero modes \fouri\  carry zero angular momentum as was
implicitly assumed in the discussion in the previous section. But isovector
fermions in the $\bf 3$ necessarily have $K \ne 0$ since they transform
in the product ${\bf 3} \times {\bf 2} = {\bf 2} + {\bf 4} $.  Since the zero
modes
for adjoint fermions are two-fold degenerate but not four-fold degenerate the
only possibility is that they carry spin $1/2$. Thus we can write
\eqn\fourt{ \psi = a_{0,1/2} \psi_0^\half + a_{0,-1/2} \psi_0^{- \half} +\hbox{non-zero modes} }
where the $\pm 1/2$ indicate the component of spin along say the $z$-axis.
Following the previous analysis we then have a four-fold degenerate
spectrum consisting of the states shown below. To simplify the notation
I have dropped the zero subscript and written $\pm$ instead of
$\pm 1/2$.
\eqn\fouru{\eqalign{ {\rm State}   \qquad  & \quad S_z \cr
               | \Omega \rangle  \qquad & \quad 0 \cr
 a_{+}^\dagger | \Omega \rangle \qquad & \quad \half \cr
 a_{-}^\dagger | \Omega \rangle  \qquad & -\half \cr
 a_{+}^\dagger a_{-}^\dagger | \Omega \rangle
  \qquad & \quad  0 \cr }}

So we see that by coupling adjoint fermions to monopoles we can give the
monopoles
spin. Remembering that one of the original problems with the Montonen-Olive
proposal
was the lack of monopole spin, this suggests that one way to cure the problem
is
to couple the monopoles to fermions in such a way as to obtain spin one
monopoles.
We will see in the next lecture that this is indeed possible.

We saw earlier that the bosonic collective coordinates of a single charge
monopole,
its location and dyon degree of freedom, could be thought of as arising from
symmetries
of the original Lagrangian which are broken by the monopole background (these
symmetries
are not broken by the vacuum, just by the monopole background).  Since
we have also
found fermion zero modes or collective coordinates it is natural to wonder
whether
they can be viewed in the same way.  In supersymmetric theories the answer is
yes,
the fermion zero modes (for charge one only) arise
due to the supersymmetries
which are unbroken in the vacuum but are broken by the monopole background.
This is discussed in the following lecture.

\subsec{Exercises for Lecture 4}

\item{E13.} Find the unitary transformation which relates the gamma
matrices used in this section to your favorite choice of gamma
matrices.
\item{E14.} For $N_f$ Dirac fermions in the doublet of $SU(2)$ we found
$N_f$ zero modes in a one monopole background which led to creation
and annihilation operators $a_0^i$, $a_0^{i \dagger}$, $i = 1,2, \cdots N_f$
obeying the anticommutation relations
\eqn\exthirt{\eqalign{\{a_0^i,a_0^j \} & = \{a_0^{i \dagger},
a_0^{j \dagger } \} = 0 \cr
\{ a_0^i,a_0^{j \dagger} \} & = \delta^{ij} }}
\itemitem{a)} Construct operators obeying the Lie algebra of $SU(N_f)$
in terms of the $a_0^i$ and $a_0^{j \dagger} $.
\itemitem{b)} Show that the monopole ground state has multiplicity
$2^{N_f}$. What representations of $SU(N_f)$ occur?
\itemitem{c)} Show that one can in fact construct generators of
$SO(2N_f)$ in terms of the $a_0^i$ and $a_0^{j \dagger} $ and that
the previous $SU(N_f)$ is embedded as $SO(2 N_f) \supset SU(N_f)$
with $2 N_f \rightarrow N_f + \bar {N_f} $ and that the monopole
ground state transforms as the (reducible) spinor representation
of $SO(2N_f)$ which
decomposes as a sum of anti-symmetric tensor representations
\eqn\exeleven{ 2^{N_f} \rto \sum_{M=0}^{N_f} \left (\matrix{
N_f\cr M\cr }\right ) . }
For assistance with this problem see \gswa.
\item{E15.} Construct the two isovector fermion zero modes $\psi_0^{\pm 1/2}$
for a charge one BPS monopole by solving the Dirac equation in this background.
Construct the operator $\vec K$ and verify that the zero modes carry angular
momentum $\pm 1/2$. For assistance see \osborn.

\newsec{Lecture 5}

\subsec{Monopoles in $N=2$ Supersymmetric Gauge Theory}
In this lecture we will be considering theories with either $N=2$ or $N=4$
spacetime supersymmetries. Realistic (i.e. chiral) models of particle
interactions
have only $N=1$ supersymmetry.  There are  theoretical reasons for
discussing $N=2$ and $N=4$, the main one being that the dynamics of these
theories is under much better control  and this allows one to make statements
about the spectrum which are valid non-perturbatively. One aspect of this
which is discussed in the next section is the fact that the Bogomol'nyi bound
follows as a consequence of the supersymmetry algebra for $N>1$. Related
to this is the fact that each supersymmetry relates field whose spin differs
by $1/2$. If we want all the fields we have discussed so far, $A_\mu, \psi,
\Phi$ with
spins ranging from $1$ to $0$, to be related by supersymmetry than we require
at least $N=2$ supersymmetry. Monopoles in $N=2$ supersymmetric
gauge theories were
first discussed in detail in reference \dadda. 

If we count bosonic and fermionic degrees of freedom for the Lagrangian
given by the sum of \threem\ and \foura\ (taking $\langle \Phi \rangle = 0$ for
the moment ) we have $3$ physical bosonic degrees for each element of the
adjoint representation of $SU(2)$ (two from the gauge fields and one from the
Higgs
field) and $4$ fermionic degrees of freedom. Thus to have the possibility of a
supersymmetric spectrum we must add an additional boson to the theory. This can
be achieved by adding another Higgs field in the adjoint representation. This
then gives the field content of $N=2$ Super Yang-Mills theory.  The Lagrangian,
in component form, is given by
\eqn\fivea{\eqalign{ {\cal L}_{N=2} = {\rm Tr}  ( & - {\quar} F_{\mu \nu}
F^{\mu \nu} - \half
(D_\mu P)^2 - \half (D_\mu S)^2 - {e^2 \over 2} [ S,P]^2  \cr
& + i {\bar \psi } \gamma^\mu D_\mu \psi
- e {\bar \psi} [ S, \psi ] - e {\bar \psi} \gamma_5 [P,\psi]  ) \cr } }
where all fields are written as elements of the Lie algebra of $SU(2)$, \ie $S
= S^a T^a$
etc. and $S$ and $P$ are scalar Higgs fields.  The Lagrangian \fivea\ is
invariant
under the supersymmetry transformations
\eqn\fiveb{\eqalign{ \delta A_\mu & =  i {\bar \alpha} \gamma_\mu \psi -
          i {\bar \psi } \gamma_\mu \alpha   \cr
        \delta P      & =    {\bar \alpha} \gamma_5 \psi -
        {\bar \psi} \gamma_5 \alpha \cr
     \delta S     & =  i {\bar \alpha } \psi -
               i {\bar \psi} \alpha  \cr
   \delta \psi  &  =   ( \sigma^{\mu \nu} F_{\mu \nu} - \Dsl S 
   +i \Dsl P \gamma_5 -i [P,S] \gamma_5 ) \alpha      \cr   }}
with $\alpha$ the Grassmann valued (Dirac) spinor supersymmetry parameter.
Since the minimal $N=1$ supersymmetry has one Majorana parameter and a Dirac
spinor is equivalent to two Majorana spinors, \fivea\ has $N=2$ supersymmetry.

There is a potential term in the Lagrangian \fivea\ but it has an exact flat
direction
whenever $[S,P]=0$. Also, as in the simpler Lagrangians we considered with
$V(\Phi) \equiv 0$, \fiveb\ is classically scale invariant and this scale
invariance
will be spontaneously broken by having a non-zero expectation value for the
scalar fields. This is enough to ensure that at least classically there will be
a massless Higgs field which is the dilaton of spontaneously broken scale
invariance.

As before, we can impose as a boundary condition that $S^a S^a \rto v^2$ as
$r \rto \infty$ thus breaking the gauge symmetry from $SU(2)$ down to $U(1)$.
It should also be clear that we can trivially obtain a charge one BPS monopole
solution in this theory using  \blatz\ with $\Phi^a$ replaced by $S^a$ 
\foot{We can always choose $P^a$ = 0 by a chiral rotation} so that
we obtain a solution obeying the Bogomol'nyi equation
\eqn\fivec{ B_i = D_i S }

Now following the earlier discussion we can ask whether this solution is
invariant
under the action of supersymmetry.  Since we are starting with a classical
solution
with the fermion fields set to zero the supersymmetry variation of the bosonic
fields
is automatically zero. The supersymmetry variation of the fermion field $\psi$
for
this background is
\eqn\fived{ \delta \psi = ( \sigma^{\mu \nu} F_{\mu \nu} - \Dsl S) \alpha }
Now using  \fivec\ and writing
\eqn\fivee{ \Gamma_5 = \gamma_0 \gamma_5 = \pmatrix{ 1 & 0 \cr 0 & -1 \cr } }
we obtain
\eqn\fivef{ \delta \psi = \gamma^i B_i (1 - \Gamma_5) \alpha }
Thus if we decompose $\alpha$ in terms of $\alpha_\pm = (1 \pm \Gamma^5)
\alpha/2 $
we see that the supersymmetries $\alpha_+$ are unbroken in the monopole
background while the $\alpha_-$ supersymmetries are broken.  The variations
\fivef\ for the broken supersymmetries give zero energy Grassmann variations
of the monopole solution, that is they are zero modes of the Dirac equation
in the monopole background as can be seen by comparing \fivef\ with
the solution of Exercise E14.

\subsec{The Bogomol'nyi Bound Revisited}
Supersymmetry also gives important new insight into the Bogomol'nyi bound \WO.
It is somewhat easier to work with the two independent Majorana components of
the
supersymmetry charge $Q_{\alpha i}$ with $\alpha$ being a spinor index and
$i=1,2$
labeling the supersymmetry. The $N=2$ supersymmetry algebra then takes
the form
\eqn\fiveg{\{ Q_{\alpha i}, {\bar Q}_{\beta j} \} = \delta_{ij}
\gamma^\mu_{\alpha \beta} P_\mu
+ \delta_{\alpha \beta} U_{ij} + (\gamma_5)_{\alpha \beta} V_{ij} }
where $U_{ij} = - U_{ji}$ and $V_{ij} = - V_{ji} $ are central terms which
commute with
the rest of the supersymmetry algebra.  They can be evaluated in a specific
theory
by constructing the supercharges in terms of the underlying fields and then
using
the canonical (anti-)commutation relations. The calculations are detailed but
straightforward and in the theory we are considering Witten and Olive found
that
\eqn\fiveh{U_{ij} = \epsilon_{ij} v Q_e, \qquad V_{ij} = \epsilon_{ij} v Q_m }
with $(Q_e,Q_m)$ the electric and magnetic charge operators described
previously.

It is then not hard to show that the supersymmetry algebra \fiveg\ implies the
Bogomol'nyi bound  $ M \ge v \sqrt{ Q_e^2 + Q_m^2} $. For example
consider the case $Q_m=0$. In the rest frame $P_\mu = (M, \vec 0 ) $
the supersymmetry algebra has the form
\eqn\fiveha{\{Q_{\alpha i}, Q_{\beta j} \} = \delta_{ij} \delta_{\alpha
\beta} M + v \epsilon_{ij} \gamma^0_{\alpha \beta} Q_e }
The left hand side is positive definite while the second term on
the right hand side has eigenvalues $\pm v Q_e$. We therefore
conclude that $M \ge v |Q_e |$.

It is clear from the above argument that the bound is saturated precisely when
one of the $Q_{\alpha i}$ is represented by zero, that is for states
annihilated
by at least one of the supersymmetry operators. This gives a beautiful relation
between  partially unbroken supersymmetry and BPS saturated states.
In fact, we can turn the argument around and derive the Bogomol'nyi equation
$B = D S$ by demanding that half of the supersymmetries \fiveb\ annihilate
the monopole solution.

There is also a close connection between BPS saturated states and short
representations of the $N=2$ supersymmetry algebra. Roughly speaking
what happens is the following. A massive representation of the
$N=2$ supersymmetry algebra is constructed by first going to
the rest frame. The supersymmetry algebra then has the same form
as a Clifford algebra and one can view linear combinations of the
supercharges as creation and annihilation operators. The smallest
representation of this algebra then has dimension $2^{2N} $ which
is $16$ for $N=2$. On the other hand for massless representations
one can go to a null frame and in this frame one finds that half
of the supersymmetry charges anticommute to zero and are thus
represented trivially. As a result  representations have
dimension $2^N = 4$. Now the $N=2$ multiplet we started with
consisting of $(A_\mu,\psi,S,P)$ has $8$ states and consists of
two irreducible massless representations of $N=2$. When we take into
account the Higgs mechanism some states eat others to get massive,
but the total number of states does not change. We thus have $8$
massive states. But this seems to contradict the previous analysis.
The resolution of this is that in constructing massive representations
the anticommutator of supercharges involves a particular combination of
the mass and central charges $V_{12}$, $U_{12}$. For a special relation
between the mass and central charges these combinations vanish and one
again must represent only $1/2$ as many supercharges non-trivially.
This special relation if of course just the Bogomol'nyi bound.
Further details on representations of $N=2$ and the role of
central charges can be found for example in \wessb.

\subsec{Monopoles in $N=4$ Supersymmetric Gauge Theory}

The Super Yang-Mills theory with $N=2$ supersymmetry we have been discussing
so far describes a single vector multiplet with physical fields $(A_\mu, \psi,
S,P)$.
One can add to this theory hypermultiplets in an arbitrary representation of
the gauge group. Hypermultiplets have a field content consisting of two
Weyl fermions and four real scalars with quantum numbers so that they
are in a real representation of the gauge group. If we consider a theory with
one vector multiplet and one hypermultiplet, both in the adjoint representation
of the gauge group and write down all possible renormalizable coupling
consistent
with $N=2$ supersymmetry then it is known that the resulting theory in fact
has $N=4$ supersymmetry.

Another more fundamental way to think about the $N=4$ theory uses
the notion of dimensional reduction from a higher dimensional
theory \bss. $N=1$ supersymmetric Yang-Mills theory with field content
consisting only of gauge fields and their supersymmetric gaugino
partners is possible only in $D=3,4,6$ and $10$ spacetime dimensions.
In order to have the correct matching of physical degrees of freedom
on must impose conditions of the fermion fields and supersymmetries.
In $D=3$ dimensions the supersymmetry and fermion fields must be
Majorana, in $D=4$ Majorana or Weyl, which are equivalent, in $D=6$
Weyl and there is no Majorana condition, and finally in $D=10$ on
must impose the Majorana and Weyl conditions simultaneously.

Thus in ten dimensions we start with a spinor $\lambda$ in the adjoint
representation of some group $G$ (which we choose to be $SU(2)$
for simplicity)  and obeying both a chirality
condition and Majorana condition:
\eqn\fiveha{(1 + \Gamma_{11} ) \lambda = 0, \qquad \bar \lambda =
\lambda^T C }
with $C$ the charge conjugation matrix. The $N=1$ Lagrangian in
ten dimensions  with $A,B = 0,1,2 \ldots 9 $ is then
\eqn\fivehb{ = {\rm Tr}  ( - {\quar} F_{A B} F^{A B}
 + { i \over 2}  {\bar \lambda } \gamma^A D_A \lambda ) }

The dimensional reduction of this Lagrangian is carried out in
detail in \osborn\ specifically for the purpose of analyzing
the monopole spectrum and the details will not be repeated here.
However some general features should be pointed out. From a group
theoretical point of view  the dimensional reduction reduces
the Lorentz group via $SO(9,1) \supset SO(3,1) \times SO(6) $.
In the dimensional reduction we discard all dependence on six of
the coordinates, as a result the $SO(6)$ part of the
ten-dimensional Lorentz group will acts as a global symmetry of
the $N=4$ theory. The gauge fields and fermion fields transform
under this reduction as
$ 10 \rightarrow  (4,1) + (1,6) $ and $16 \rightarrow (2_+, 4_+)
+(2_-, 4_-) $ respectively where the subscript indicate the
chirality. As a result the four-dimensional spectrum will consist of
a gauge field, six scalars $\Phi_a$ in the $6$ of $SO(6)$ and
four Weyl spinors $\lambda_i$
transforming as a $4$ of $SO(6)$ (or to be precise
$Spin(6) \equiv SU(4)$ ).  The resulting Lagrangian can be written
in a variety of forms, not all of which make the $SO(6)$ symmetry
manifest. Probably the most elegant formalism uses the fact that
the $6$ of $SU(4)$ arises in the antisymmetric product $(4 \times 4)_A$
to write the six scalar fields in terms of an antisymmetric complex
matrix $\Phi_{ij}$, $i,j = 1 \ldots 4$ obeying the condition
$(\Phi_{ij})^\dagger = \Phi^{ij} = (1/2) \epsilon^{ijkl} \Phi_{kl} $.
The Lagrangian
is then in two component form
\eqn\fivehc{\eqalign{ {\cal L}_{N=4} = &{\rm Tr}
( - {\quar} F_{\mu \nu} F^{\mu \nu}
+ i \lambda_i \sigma^\mu D_\mu \bar \lambda^i + \half D_\mu \Phi_{ij}
D^\mu \Phi^{ij} \cr & + i \lambda_i [ \lambda_j , \Phi^{ij} ]
+ i \bar \lambda^i [ \bar \lambda^j , \Phi_{ij} ] + {\quar}
[\Phi_{ij}, \Phi_{kl} ] [ \Phi^{ij} , \Phi^{kl} ] ) \cr }}

As in our discussion of the $N=2$ theory, it is clear that the $N=4$
theory is classically scale invariant, that the potential
has flat directions, and that we can give expectation values to
the scalars which will break the $SU(2)$ gauge symmetry to $U(1)$
and spontaneously break scale invariance. For example a simple choice
would be to take only the field $\Phi_{12} $ to have a non-zero expectation
value. We could then embed the BPS solution into this theory by
replacing the Higgs field $\Phi$ in the BPS solution by $\Phi_{12}$.

Giving a non-zero expectation  value to $\Phi_{12}$ not only
breaks the $SU(2)$
gauge symmetry  to $U(1)$, it also breaks the classical scale invariance
and also spontaneously breaks the global $SO(6)$ symmetry to $SO(5)$.
It is known that $N=4$ Super Yang-Mills theory is a finite theory
with vanishing beta function and thus an exact quantum scale invariance.
The scalar spectrum after gauge symmetry breaking will therefor consist
of one massless scalar (the dilaton) which is the Nambu-Goldstone (NG)
boson of spontaneously broken scale invariance and five massless
scalars which are the NG bosons of the spontaneously broken $SO(6)$
global symmetry.

{}From \fivehc\ we see that the fermions have the standard Yukawa and
gauge couplings to the fields appearing in the BPS monopole solution.
It should thus
be clear that the charge one BPS monopole embedded in the $N=4$ theory
has twice as many fermion zero modes as we had for the pure $N=2$ theory since
there are now the equivalent of two Dirac fermions (i.e. four Weyl
fermions) in the adjoint representation
of $SU(2)$. Following the discussion around \fourt\ we will now have fermion
zero
modes in the one monopole sector  $a_{0 \pm1/2}^n $with $n=1,2$ labeling the
two fermion fields. Dropping the zero subscript and writing $\pm$
for $\pm 1/2$ as before the spectrum will
then consist of the states
\eqn\fiveclam{\eqalign{ {\rm State}   \qquad  & \quad S_z \cr
                                  | \Omega \rangle  \qquad & \quad 0 \cr
     {a_{\pm}^n}^\dagger | \Omega \rangle \qquad & \pm \half \cr
    {a_{-}^n}^\dagger {a_{+}^m}^\dagger | \Omega \rangle
          \qquad  & \quad 0 \cr
    {a_{+}^1}^\dagger {a_{+}^2}^\dagger | \Omega \rangle
           \qquad & \quad 1 \cr
    {a_{-}^1}^\dagger {a_{-}^2}^\dagger | \Omega \rangle
        \qquad & -1 \cr
   {a_{\mp}^1}^\dagger {a_{\mp}^2}^\dagger
    {a_{\pm}^n}^\dagger
    | \Omega \rangle  \qquad & \mp \half \cr
  {a_{+}^1}^\dagger {a_{+}^2}^\dagger  {a_{-}^1}^\dagger
    {a_{-}^2}^\dagger \Omega \rangle  \qquad & \quad 0 \cr }}
for a total of $16$ states, $8$ bosons, $6$ with spin $0$ and $2$ with spin
$\pm 1$
and $8$ fermions with spin $\pm 1/2$. This is the same as the
content of the gauge super multiplet
of $N=4$ Yang-Mills theory.

Thus in $N=4 $ gauge theory we finally see that it is possible to obtain
monopoles
of spin one and in fact the monopole supermultiplet and the gauge
supermultiplet
are the same in this theory. This in fact is not so surprising, there is a
unique multiplet
in $N=4$ gauge theory which does not contain spin greater than one.

\subsec{Supersymmetric Quantum Mechanics on ${\cal M}_k$ }

We argued earlier that we can think of the fermion zero modes
as Grassmann collective coordinate for the monopole moduli space.
In the absence of fermion fields the collective coordinate expansion
with background fields $A^i(x^j, z^\alpha(t))$, $\Phi(x^j, z^\alpha(t))$
leads to a low-energy effective action by substituting into the
four-dimensional action and integrating over $R^3$ to obtain
\eqn\fiveseff{S_{\rm eff} = \int dt {\cal G}_{\alpha \beta} \dot z^\alpha
\dot z^\beta }
with ${\cal G}$ the metric on the monopole moduli space. In other words,
at very low energies the field theory about the monopole background
can excite only a finite number of degrees of freedom, those that
correspond to motion along the moduli space, and thus we can reduce
the dynamics to quantum mechanics.

We would like to add fermion zero modes to this picture. We can do this
by also expanding the fermion fields as
\eqn\fiveferm{\psi = \sum_{\alpha} \lambda_\alpha(t) \psi_{0 \alpha}(x,z) }
where the $\psi_{0 \alpha}$ are the fermion zero modes, treated as
real numbers, and the $\lambda_\alpha(t)$ are Grassmann valued fermion
collective coordinates. One can include these in the quantum mechanical
effective action by carrying out the same procedure as before. The
details are somewhat subtle however and will not be presented
here. For details see \gauntwo\ for the analysis in $N=2$ theories
and \blumfour\ in $N=4$ theories.  However the answer is not surprising.
The quantum mechanical action is extended to the action for a
supersymmetric quantum mechanics. In the $N=2$ case there are four real
supersymmetries in spacetime which are unbroken in the monopole
background. As a result we expect an action with $N=4$ world-line
supersymmetry. This action is
\eqn\fiveard{ S_{eff} = \half \int dt {\cal G}_{\alpha \beta}
( {\dot z}^\alpha {\dot z}^\beta +
4 i {\lambda^\dagger}^\alpha D_t \lambda^\beta ) + {\rm const} . }
where
\eqn\covderdef{
  D_t \lambda^\alpha = { d \lambda^\alpha \over dt } +
  \Gamma^{\alpha}_{\beta \gamma} { dz^\beta \over dt} \lambda^\gamma }
is the covariant derivative acting on the spinor $\lambda$.
The nomenclature for supersymmetry in quantum mechanics is a bit
confusing. Originally actions with two component spinors and one
supersymmetry were constructed and the supersymmetry was referred to
as $N=1$. It was later realized that one could also have one supersymmetry
with one component spinors and this was unfortunately called
$N=1/2$ supersymmetry. With this nomenclature the action \fiveard\ might
be said to possess $N=4 \times 1/2 $ supersymmetry. The presence of
four supersymmetries requires that the moduli space be a
hyperkahler manifold.
This means ${\cal M}_k$ has three complex structures ${\cal J}^m $
which obey
\eqn\fivej{ {{ \cal J}^m_\beta}^\alpha {{\cal J}^n_\alpha}^\gamma = -
\delta_\beta^\gamma
\delta^{m n} + \epsilon^{mnp} {{\cal J}^p_\beta}^\gamma }
and the action has $N=4 \times 1/2$ supersymmetry with supersymmetry
transformations
\eqn\fivek{\eqalign{ \delta z^\alpha = & i \beta_4 \lambda^\alpha + i \beta_m
\lambda^\beta
{{\cal J}^m_\beta}^\alpha \cr
\delta \lambda^\alpha = & - {\dot z}^\alpha \beta_4 - \beta_m {\dot z}^\beta
 {{\cal J}^m_\beta}^\alpha . \cr }}
The BPS monopole moduli space can be shown to be hyperkahler, independent
of supersymmetry \ah, but to a physicist, supersymmetry provides the
simplest explanation  of this fact.

If we choose one of the complex structures to introduce complex coordinates on
${\cal M}_k$ then we have the canonical anti-commutation relations
\eqn\fivel{ \{ \lambda^\alpha , \lambda^{\bar \beta} \} = \delta^{\alpha \bar
\beta} .}
The $\lambda^{\bar \alpha}$ therefore act as creation operators and
we thus have a spectrum of
states of the form
\eqn\fivem{ f_{\bar \alpha_1 \cdots \bar \alpha_p} \lambda^{\bar \alpha_1}
\cdots
\lambda^{\bar \alpha_p} | \Omega \rangle .}
These
state are in one to one correspondence with
holomorphic $(0,p)$ forms on ${\cal M}_k$
\eqn\fiven{ | f \rangle = f_{\bar \alpha_1 \cdots \bar \alpha_p} \lambda^{\bar
\alpha_1} \cdots
\lambda^{\bar \alpha_p} | \Omega \rangle  \Longleftrightarrow
f_{\bar \alpha_1 \cdots \bar \alpha_p}  d z^{\bar \alpha_1} \wedge \cdots d
z^{\bar \alpha_p} .}

In the reduction of $N=4$ Super Yang-Mills theory one obtains a Lagrangian
with  $N=4 \times 1$ supersymmetry (twice as much supersymmetry as
the $N=2$ theory) given by
\eqn\fiveo{ S_{eff} = \half \int dt  {\cal G}_{\alpha \beta} 
( {\dot z}^\alpha {\dot z}^\beta
+ i {\bar \psi}^\alpha \gamma^0 D_t \psi^\beta ) + {1 \over 6} R_{\alpha \beta
\gamma \delta}
( {\bar \psi}^\alpha \psi^\gamma ) ( {\bar \psi}^\beta \psi^\delta ) }
where now $\psi^\alpha$ is a two-component spinor rather than a one-component
object as in \fiveard. As a result of this doubling one now finds that the
Hilbert space
of states is the same as the space of all differential forms on ${\cal M}_k$
and that
the Hamiltonian is the Laplacian acting on forms. For further details
of this correspondence see \refs{\wittsusy, \smon}.

\subsec{Exercises for Lecture 5}
\item{E16.} Verify that $\delta \chi = \gamma^i B_i \alpha_- $ is a zero-mode
of the Dirac equation and that this
agrees with what you found in Exercise 12.
\item{E17.} Construct the supercharges $Q$, $Q^*$ for the theory described
by \fiveard. Show that with the above correspondence between states and forms
the supercharges and Hamiltonian are given by
\eqn\exfourteen{\eqalign{ Q^* & \Leftrightarrow \bar \partial \cr
                                                 Q & \Leftrightarrow {\bar
\partial}^\dagger \cr
              H = \{ Q, Q^* \}  & \Leftrightarrow {\bar \partial}^\dagger {\bar
\partial} +
           {\bar \partial } {\bar \partial }^\dagger = \half ( d d^\dagger +
d^\dagger d ) }}
where the latter equality uses the fact the ${\cal M}_k$ is Kahler. Thus the
Hamiltonian is just the Laplacian acting on forms.
\item{E18.} Show that the $N=2$ Yang-Mills action \fivea\ can be derived
by dimensional reduction of $N=1$ Yang-Mills theory in six
spacetime dimensions.

\newsec{Lecture 6}

\subsec{Implications of $S$-duality}
It is time to take stock of where we are in the search for theories which may
exhibit
an exact electromagnetic duality. We have seen that this cannot be the case in
pure $SO(3)$ gauge theory or in $N=2$ Yang-Mills theory because the monopole
does not have spin one and thus cannot be dual to the
$W$ boson \foot{This does not
rule out a duality relating monopoles to fermion matter fields in $N=2$
theories, strong
evidence for such
a duality was found in \swtwo\ but a full discussion of this would lead us to
far afield.}.
On the other hand in $N=4$ Yang-Mills theory there is only one supermultiplet
which contains only spin $\le 1$ and we have seen that the monopoles and gauge
bosons both lie in this supermultiplet.

In addition, although we will not discuss it in these lectures, quantum
corrections
in $N=4$ are under very precise control. In  fact it is known that this theory
has
vanishing beta function, both perturbatively and non-perturbatively. This
ensures
that the flat direction in the potential we are utilizing remains  in the full
theory
and that BPS states constructed at weak coupling continue to exist and evolve
smoothly to states at strong coupling.

We have thus addressed the first two objections to the Montonen-Olive proposal.
We are still faced  with finding a non-trivial way to check the proposed
duality without having to compute directly at strong coupling. This is where
the extension of  duality to $SL(2,Z)$ plays a central role as was
first appreciated by Sen.  We will make one assumption, namely that
the state with $(n_e,n_m)=(1,0)$  (the $W^+$ boson) exists at all values
of the coupling $\tau$ with a degeneracy of $16$ corresponding to the $16$
states in the short vector representation of $N=4$ supersymmetry.  This
is an extremely mild assumption. We have already argued that
the dimension of the representation cannot change as parameters of the
theory are varied and we know that such a state exists at weak coupling
as a BPS saturated state. The one known mechanism by which BPS saturated
states can disappear requires that the lattice spanned by the
electric and magnetic charges degenerate
at some value of $\tau$ and in this theory that is ruled out
by the non-renormalization theorems.

Given the existence of the $(1,0)$ state $SL(2,Z)$ duality requires the
existence of all the $SL(2,Z)$ images of this state. Since a $SL(2,Z)$
transformation acts on this state as
\eqn\sixa{ \left (\matrix{ 1\cr 0\cr }\right ) \rto \pmatrix{ a & b \cr c & d
\cr } \left (\matrix{ 1\cr 0\cr }\right )
 = \left (\matrix{ a\cr c\cr }\right ) }
duality requires the existence of states with $(n_e,n_m)= (a,c)$ with the
same degeneracy of $16$. Furthermore, since $ad-bc =1$ it follows that
$a$ and $c$ are relatively prime, $(a,c)=1$.  Also, by starting at a value
of $\tau$ corresponding to strong coupling and then performing the
duality transformation \sixa\ these states must exist at weak coupling and
thus should appear in a semi-classical analysis of the spectrum.

For $c=1$ we require states $(a,1)$ for arbitrary integer $a$. These are just
the dyonic excitations of the single charge BPS monopole and from
the previous analysis we see that such states do exist with the correct
multiplicity. Furthermore these states are BPS states as is
demonstrated in \gauntwo.  For $c=2$ we require states $(a,2)$ with $a$ odd,
again with
a degeneracy of $16$.   Previous to Sen's analysis such states were not known
to exist.

Constructing these states in the full $N=4$ supersymmetric field
theory would be very difficult. Luckily the question can be reduced
to construction of bound states in the moduli space approximation.
To see why this is the case
consider a bound state with electric charge one. A BPS monopole
state with $(n_e,,n_m)= (0,1)$ has mass $M_{(0,1)} = vg $ while
a dyon state with charge $(1,1)$ has mass $M_{(1,1)}= v \sqrt{e^2 + g^2}$.
A BPS bound state of charge $(2,1)$ on the other hand has mass
$M_{(2,1)} = v \sqrt{4g^2 + e^2}$. At weak coupling the binding energy
is thus
\eqn\sixgab{M_{(2,1)} - M_{(1,1)} - M_{(1,0)} \sim ve (e/4g) << ve .}
Since this is much less than the $W$ mass we should be able to study
the existence of this bound state in the moduli space approximation.
The same argument applies to states of greater electric charge at
sufficiently weak coupling.

In the following we will see how
the existence of these states follows from
a careful analysis of supersymmetric quantum mechanics on ${\cal M}_2$.
Evidence for the existence of states with arbitrary $(a,c)$=1 can be found
in \refs{\segal,\porrati}. What about states with charges $(0,n_m)$ or
$(n_e,0)$ ? At weak coupling we know that there are no electric charge
two bound states in the spectrum, there are BPS states in the theory with
charge $(n_e,0)$ but these are not distinct from the multi particle
continuum of states. Similarly, although there are charge $(0,n_m)$
monopoles, our analysis will show (for $n_m=2$)
that these are not normalizable
bound states of $n_m$ single charge monopoles but just part of
the continuum of states of $n_m$ single charge monopoles \foot{This
is in distinction to the situation for fundamental strings or some
string, D-brane configurations where there are discrete states with
multiple charge which should thought of as bound states and distinguished
from the multiparticle continuum \dab.}.

\subsec{The Two-monopole Moduli Space From Afar}
As we will see in the following section, the metric on the two-monopole moduli
space can be constructed exactly and the spectrum of supersymmetric quantum
mechanics on this  space can thus be determined by explicit calculations.
However instead of proceeding directly to this analysis I would like
to discuss briefly a description of the asymptotic form of the
two monopole moduli space due to Manton \mantonmod. There are two
reasons for doing this. First, it brings out the physics of the
moduli space in a direct way that is not obscured by difficult
mathematics or special functions. Second, this approximate description
has played a role in studies of the multi-monopole
moduli space \gibbmannew\
and other problems involving moduli spaces.

We begin with the fact that the BPS monopole has magnetic charge
and dilaton charge both equal to $g$ with the definition of dilaton
charge given in \twofc. At large distances from the monopole we
can summarize this fact by writing down a point like interaction between
the monopole and the photon and dilaton fields with equal strength
interactions. We begin with the interaction with the photon field.
Instead of working with the conventional vector potential $A_\mu$
as in the previous sections it is useful to introduce a dual potential
$\tilde A^\mu = (\tilde A^0, \tilde {\vec A } )$
defined by $\tilde F = d \tilde A $ in order to
describe the field of a point monopole. We can then couple a point
monopole of mass $M$ to the photon by mimicking the interaction of a
electrically charged point particle \foot{The form of the action
below is not manifestly covariant since it involves the time $t$
as a parameter rather than the proper time $\tau$. It does of course
lead to the correct covariant equations of motion \golds\ and is more
convenient for our purposes.}:
\eqn\sixb{ S_{\tilde A} = \int dt  \left( - M \sqrt{1 - {\vec v}^2}
- g \tilde A^0
+ g {\vec v} \cdot \tilde {\vec A } \right) . }
When this is coupled to the electromagnetic action
\eqn\sixba{ S_{EM} = - {1 \over 4} \int d^4 x F_{\mu \nu} F^{\mu \nu}
= - {1 \over 4} \int d^4 x {\tilde F}_{\mu \nu} {\tilde F}^{\mu \nu} }
one finds as required that a monopole at rest at the origin gives
rise to a Coulomb magnetic field, and the action of this field on
a second monopole gives rise to the standard Coulomb repulsion between
like sign monopoles.  Now since the theory also includes a massless
dilaton field with action
\eqn\sixbb{S_{dil} = {v^2 \over 2} \int d^4x \partial_\mu {\cal D}
\partial^\mu {\cal D} }
we must also include the coupling to the dilaton to obtain the
correct force law. The coupling to the dilaton is dictated by
the fact that a shift of the dilaton is equivalent to a shift in the
mass of the monopole. This is the statement of spontaneously broken
scale invariance. We can thus generalize \sixb\ to
\eqn\sixbc{ S_{\tilde A, {\cal D}} = \int dt
\left(  ( - M  + v {\cal D})  \sqrt{1 - {\vec v}^2} - g \tilde A^0
+ g {\vec v} \cdot \vec {\tilde A }  \right) }
and now if we compute the net force between two stationary monopoles
we find that the Coulomb repulsion is precisely cancelled by the
dilaton attraction.

Now we can ask what happens if the monopoles move relative to each other
at low velocities. At small velocities and at large impact parameter
the interactions will still be mediated by exchange of massless particles
so the previous description should suffice. If the first
monopole is moving at velocity
$\vec v_1 << 1$  a standard computation of the
Lienard-Wiechert potentials  and dilaton field
to first order in the velocity gives
\eqn\sixbd{\eqalign{ {\tilde A}_0 & = {g \over 4 \pi r }, \cr
              {\vec {\tilde A}} & = {g \over 4 \pi r} \vec v_1 ,\cr
             v {\cal D} & ={g \over 4 \pi r } \sqrt{1- {\vec v}_1^2} ,\cr }}
with $\vec r$ the relative separation between the monopoles.
If we now substitute these fields into the Lagrangian for the second
monopole and separate out the center of mass motion we are left
with an action which governs the relative motion of the
two monopoles:
\eqn\sixbe{ S_{rel} = \int dt ( {M \over 4} - {g^2 \over 8 \pi r} )
{d \vec r \over dt} \cdot {d \vec r \over dt } . }

The action \sixbe\ has no potential term, and a spatially varying kinetic
term.  We can interpret it as saying that the relative motion is geodesic
motion for a metric  on $R^3$ given by $ds^2 = U(r) d \vec r \cdot d \vec r$
with $U(r) = 1-g^2/(2 \pi M r)$. Note that so far everything we have
done could also have been done for a electrically charged $W$ boson
in this theory, the low velocity motion of purely electrically or
magnetically charged particles in this theory is equivalent to geodesic
motion. Physically, the forces due to
photon and dilaton
exchange no longer cancel at non-zero velocity due to the different
retardation effects for spin zero and spin one exchange.

Now for monopoles there is a natural generalization of this result.
We saw previously that the classical single monopole moduli space has the
form $R^3 \times S^1$ where the ``velocity'' on the $S^1$ factor
determines the dyons electrical charge. Thus to generalize the
above result to scattering of dyons we should include the electrical
charge of the dyons and view this as a velocity in some additional
coordinate on $S^1$. The analysis is slightly complicated by the fact
that one must use both $A$ and $\tilde A$ in the computation, but is
essentially a straightforward generalization of what we have done.
In carrying out the computation one should remember that the dilaton
must couple to $\sqrt{e^2 + g^2}$ since this is the mass of the
dyon from the BPS bound.
The analysis is carried out in \mantonmod\ with the result that the
relative motion of two dyons with electric and magnetic charges
$(e_1,g_1)$, $(e_2, g_2)$ and relative electric charge $e=e_2-e_1$
is given by
\eqn\sixbf{ S_{rel} = \int dt ( {M \over 4} - {g^2 \over 8 \pi r} )
{d \vec r \over dt} \cdot {d \vec r \over dt } +{eg \over 4 \pi}
{d \vec r \over d t} \cdot \vec \omega + {e^2 \over 8 \pi r } }
where $\vec \omega $ is the Dirac monopole potential determined
by $\vec \nabla \times \omega = \vec r /r^3 $.  From the previous
discussion we would like to regard the electric charge $e$ as the
velocity along a $S^1$ governing the relative charge of the
two monopoles. If we call this fourth relative coordinate $\chi$
with $e \sim \dot \chi $ then the equations of
motion following from \sixbf\ are equivalent to the equations for
geodesic motion in the metric
\eqn\sixbg{ ds^2 = U(r) d \vec r \cdot d \vec r + {g^2 \over 2 \pi M U(r)}
(d \chi + \vec \omega \cdot d \vec r )^2 . }
The metric \sixbg\ is known in the relativity literature as the
Taub-NUT metric with negative mass.  From our derivation, we only
expect it to agree with the exact metric on the two monopole
moduli space for monopole separations large compared to the
inverse $W$ mass where the Dirac monopole approximation is valid.

\subsec{The Exact Two-monopole Moduli Space}

Although the previous analysis gives a nice physical picture of
the metric on the two monopole moduli space at large separation,
we need the full metric in order to provide a precise test
of duality.
The exact two-monopole moduli space has been determined by Atiyah and Hitchin
using the fact that it has $SO(3)$ isometry arising from rotational invariance,
the fact that in four dimensions hyperkahler implies self-dual curvature, and
the fact that the metric is known to be complete. This reduces
the problem to an analysis of specific differential equations which can then be
solved
in terms of elliptic functions. Luckily we will not need to investigate the
detailed form
of the metric. A good reference for what follows is \gibbman.

The two monopole moduli space has the form
\eqn\sixc{ {\cal M}_2 = R^3 \times \left( {S^1 \times {\cal M}_2^0 \over Z_2 }
\right) }
where the $R^3$ factor is the overall center of mass of the system and the
$S^1$
factor describes the overall dyon rotator degree of freedom. The reduced moduli
space ${\cal M}_2^0$ is four-dimensional and when the two monopoles are far
apart one can think of the coordinate on ${\cal M}_2^0$ as being the relative
separation of the monopoles and the relative orientation of the dyon degrees of
freedom.  Thus in this asymptotic region we have
\eqn\sixd{\eqalign{  \chi & = \half ( \chi_1 + \chi_2 ) \cr
                                     \vec X & = \half ( \vec x_1 + \vec x_2 )
\cr } }
as coordinates on $S^1 \times R^3$ and
\eqn\sixe{\eqalign{ \psi & = \half ( \chi_1 - \chi_2 ) \cr
                                   \vec x & = \half ( \vec x_1 - \vec x_2 ) \cr
} }
as coordinates on ${\cal M}_2^0$ with $(\vec x_1, \chi_1)$ the collective
coordinates
of monopole one and similarly for monopole two.  The $Z_2$ identification in
\sixc\ arises
because a $2 \pi$ rotation of one of the dyon degrees of freedom leads to the
same
monopole configuration. Explicitly it is given by the transformation
\eqn\sixf{ I_1:\quad \psi \rto \psi + \pi, \quad \chi \rto \chi + \pi .}

The explicit metric on ${\cal M}_2^0$ is given by
\eqn\sixg{ ds^2 = f(r)^2 dr^2 + a(r)^2 {(\sigma_1^R)}^2 +
b(r)^2 {(\sigma_2^R)}^2 + c(r)^2 {(\sigma_3^R)}^2 }
where the $\sigma_i$ are left-invariant one-forms on
$SO(3)= S^3/Z_2$\foot{The metric \sixg\ thus has $SO(3)$ isometry since it
is invariant under the (left) action of $SO(3)$. It may seem odd at
first sight that the metric has $SO(3)$ isometry is spite of having
different radial functions multiplying each of the terms in \sixg. However
if all radial functions were equal the metric would have $SO(4)=SU(2) \times
SU(2) \sim SO(3) \times SO(3) $ isometry given by both the left and right
actions of $SO(3)$.}.
In one particular basis they are given by
\eqn\sixga{\eqalign{\sigma^R_1&= -\sin\psi d\theta+\cos\psi \sin\theta d\phi\cr
\sigma^R_2&= \cos\psi d\theta+\sin\psi \sin\theta d\phi\cr
\sigma^R_3&= d\psi+\cos\theta d\phi\cr} }
with with $0 \le \theta \le \pi$, $0 \le \phi \le 2 \pi$, $0 \le \psi < 2
\pi$.
The angles are further restricted under the identification
of the discrete right isometry \gibbman\
\eqn\bird{
(\phi,\theta,\psi)I_x= (\pi+\phi,\pi-\theta,-\psi)
.}
Note that we can equivalently let the range of $\psi$ be $0 \le \psi
< 4
\pi$ and then divide out by $I_x$.

We will follow
\gibbman\ in choosing $f(r)= -b(r)/r$.
The radial functions $a(r)$, $b(r)$ and $c(r)$ are given
explicitly in \ah. Here we only need the asymptotic forms.
Near
$r=\pi$ they take the form
\eqn\asymp{\eqalign{
a(r)&=2(r-\pi)\left\{1-{1\over 4\pi}(r-\pi)\right\}+\dots\cr
b(r)&=\pi\left\{1+{1\over 2\pi}(r-\pi)\right\}+\dots\cr
c(r)&=-\pi\left\{1-{1\over
2\pi}(r-\pi)\right\}+\dots.\cr}}
Introducing appropriate Euler angles, it can be shown that after the
identification by $I_x$ the metric
is smooth near $r=\pi$ and that $r=\pi$ is an $S^2$ or bolt \gibbman.
Near infinity, $r\to\infty$, the functions take the form
\eqn\asympt{\eqalign{
a(r)&=r\left(1-{2\over r}\right)^{1/2}+\dots\cr
b(r)&=r\left(1-{2\over r}\right)^{1/2}+\dots\cr
c(r)&=-2\left(1-{2\over r}\right)^{-1/2}+\dots,\cr}}
where the neglected terms fall off exponentially with $r$.
It can be shown that this asymptotic metric is equivalent to
the Taub-NUT metric \sixbg.

\subsec{Duality and Sen's Two-form}

We now want to use the metric on the two monopole moduli space
to partially test the predictions of $S$ duality following
Sen's original analysis. In particular, $S$ duality predicts
the existence of BPS saturated bound states with magnetic
charge $2$ and odd electric charge.

In the moduli space approximation BPS states are supersymmetric ground states,
and these in term correspond to harmonic forms on the moduli space
as  discussed at the end of the previous lecture. The fact that
we are looking for a bound state means that the wave function or form
on the relative moduli space must be normalizable, that is in $L^2$.

In quantizing the theory on ${\cal M}_2$ we will obtain $16$ fold degenerate
states from the
wedge product of the $16$
harmonic (constant) forms on $R^3 \times S^1$ with $L^2$ harmonic forms on
${\cal M}_2^0$.
$S$-duality predicts that we have precisely this degeneracy so it requires the
existence
of a unique $L^2$ harmonic form on ${\cal M}_2^0$. Furthermore, for
the corresponding states to have odd electric charge this form must be odd
under the $Z_2$ action \sixf\ . Now since the Hodge dual of a
harmonic form is also harmonic, it follows that we can get a unique harmonic
form only if
the form is self-dual or anti-self-dual.

With this information it is then straightforward to write down
the candidate form. It is given by the ansatz
\eqn\sixh{ \omega = F(r) ( d \sigma_1 - {f a \over bc} dr \wedge \sigma_1 ).}
Note that this is anti-self-dual by construction.
Demanding that $\omega$ be harmonic yields the equation
\eqn\sixi{ {d F \over d r} = - {fa \over bc} F .}
An analysis of this equation at infinity and at the ``bolt'' shows
that the form $\omega $ is normalizable and well behaved at the
bolt \sendual.
Furthermore this is the unique such form.
This thus establishes the existence of precisely the BPS bound states
which are required by $S$ duality in the two monopole sector.

This analysis also shows that there are no such BPS bound states
in $N=2$ super Yang-Mills theory without matter \smon.
In the $N=2$ theory
supersymmetric ground states are holomorphic forms on the moduli
space, but the form \sixh\ being anti-self-dual is a $(1,1)$ form
and thus not holomorphic. Of course in the $N=2$ theory there was
no reason to expect such bound states since the theory is
not $S$ dual.

We have shown that such states exist mathematically,
but one might wonder what the physics
is behind this result and whether there is some simple explanation
why we found bound states for $N=4$ but not for $N=2$.
I believe the answer has to do with spin
dependent forces. Once the monopoles have spin there will be additional
long range forces (e.g. spin-orbit and spin-spin) besides those
considered in Manton's analysis of the asymptotic moduli space.
Depending on the magnitude of the spin these
spin dependent forces can lead to bound states which would not
otherwise exist. So for example, without supersymmetry,  bound states
of the basic Prasad-Sommerfield theory with vanishing potential would
correspond to $L^2$ harmonic functions rather than forms on the two
monopole moduli space and we have seen that these do not exist.
With $N=2$ supersymmetry the absence of a four fermion term in the
supersymmetric quantum mechanics indicates a cancellation of the
spin-spin forces between vector and scalar exchange. It is only
when we get to $N=4$ supersymmetry and spin one monopoles that
the spin dependent forces can lead to new BPS saturated bound states.

\subsec{Exercises for Lecture 6}
\item{E19.} Analyze the asymptotic form of the two monopole moduli space
given the asymptotic formulae for $a(r)$, $b(r)$, $c(r)$ and $f(r)$.
Show that the asymptotic metric can be put in the from of
the Taub-NUT metric \sixbg.
\item{E20.} The following problem is an extended exercise in the
geometry of the $SU(2)$ group manifold, a.k.a. $S^3$. Parameterize the
three sphere by Euler angles and write a general $SU(2)$ rotation
as
\eqn\sthree{
\eqalign{U(\phi,\theta,\psi)=&U_z(\phi) U_y(\theta) U_z(\psi)\cr
&=\left(\matrix{\cos{\theta\over 2}e^{i {(\psi+\phi)\over 2}}&
         \sin{\theta\over 2}e^{-i {(\psi-\phi)\over 2}}\cr
        -\sin{\theta\over 2}e^{i {(\psi-\phi)\over 2}}&
         \cos{\theta\over 2}e^{-i {(\psi+\phi)\over 2}}\cr}\right)
}}
\itemitem{a)}By expanding the one-form $U^{-1}dU $ in the
basis $ i \tau_i/2$ with $\tau_i$ the Pauli matrices construct the
left-invariant or ``right'' one-forms $\sigma_i^R$. Similarly,
expand $dU U^{-1}$ to obtain a set of right-invariant ``left''
one-forms $\sigma_i^L$. The
one-forms $\sigma_i^{R,L}$ are dual to
left (right)
invariant vector fields $\xi^R_i$ ($\xi^L_i$)
which generate right (left) group actions.
\itemitem{b)}
Construct explicitly  the dual vector
fields satisfying $\langle\xi^R_i,\sigma^R_j\rangle=\delta_{ij}$ and
$\langle\xi^L_i,\sigma^L_j\rangle=\delta_{ij}$ and show
that they are given by
\eqn\livf{
\eqalign{
\xi^R_1&= -\cot\theta \cos\psi \del_\psi-\sin\psi \del_\theta+
{\cos\psi\over \sin\theta}
\del_\phi\cr
\xi^R_2&= -\cot\theta \sin\psi \del_\psi+\cos\psi
\del_\theta+{\sin\psi\over
\sin\theta}
\del_\phi\cr
\xi^R_3&= \del_\psi\cr}}
and
\eqn\riof{
\eqalign{
\xi^L_1&= -{\cos\phi\over \sin\theta}
\del_\psi +\sin\phi \del_\theta+\cot\theta \cos\phi\del_\phi\cr
\xi^L_2&= {\sin\phi\over \sin\theta}
\del_\psi+\cos\phi \del_\theta-\cot\theta \sin\phi \del_\phi\cr
\xi^L_3&= \del_\phi.\cr}}
\itemitem{c)}
Show that the left and right invariant one forms satisfy the Maurer-Cartan
equations
\eqn\mc{
\eqalign{
d\sigma^R_i&={1\over 2}\epsilon_{ijk}\sigma^R_j\wedge\sigma^R_k\cr
d\sigma^L_i&=-{1\over 2}\epsilon_{ijk}\sigma^L_j\wedge\sigma^L_k.\cr
}}
and that the Lie brackets of the left and right vector fields are given by
\eqn\lb{
\eqalign{[\xi^R_i,\xi^R_j]&=-\epsilon_{ijk}\xi^R_k\cr
[\xi^L_i,\xi^L_j]&=\epsilon_{ijk}\xi^L_k\cr
[\xi^R_i,\xi^L_j]&=0.\cr}}
The last equation expresses the fact that the right (left) vector
fields
are left (right) invariant.

\newsec{Conclusions and Outlook}

In these lectures we have developed some of the basic tools needed
to study duality in gauge theories with extended supersymmetry and
have verified one non-trivial prediction of $S$ duality. Of course
this is a far cry from having a complete understanding of duality
or even from testing it in a comprehensive way. At the time of writing
this final section (March 1996) duality has turned into an enormous
enterprise which is changing the way that we think about both field
theory and string theory. I could not possibly summarize the current
situation or the open problems and any attempt to do so would be
obsolete within weeks. Instead let me end by mentioning some of the
progress and open problems in the much more narrowly defined area of
exact duality in $N=4$ and finite $N=2$
super Yang-Mills theory which has been the
end of the logical development of these lectures.

\item{1.}
There have been attempts to extend Sen's result to the
full set of dyons states required by $S$ duality
\refs{\segal, \porrati} but my impression is that no completely
convincing construction yet exists.
\item{2.}
All of the current tests of duality are really tests of the
BPS spectrum of states or equivalently of states
preserving half of the supersymmetry (this is true also of tests
relying on topological field
theory constructions). Yet if there is exact duality then it
must relate all states and
correlation functions including those at non-zero momentum.
So far we do not have the tools to explore duality in a dynamical
setting. For one partially successful attempt in this direction
see \hms.
\item{3.}
As mentioned at the end of lecture 4, finite $N=2$ theories are also
conjectured to exhibit an exact duality symmetry. The simplest
case involves gauge group $SU(2)$ with $N_f=4$ hypermultiplets
in the doublet representation of $SU(2)$. Some of the predictions
of duality in this theory were explored in \refs{\gaunharv, \sethi}.
\item{4.}
In these lectures I have only considered duality with gauge
group $SU(2)$. For larger gauge groups there are analogous
predictions both for $N=4$ Yang-Mills theory and for $N=2$
theories with vanishing beta function. There has been recent
progress in testing duality in these theories \refs{\newdual}.
\item{5.}
Finally, the central question remains of {\it why} these theories
exhibit duality. It many cases it
appears that duality in these theories
is a low-energy manifestation of  duality symmetries in string
theory. The origin of duality in string theory is still an
unsolved mystery.
\listrefs
\bye